\documentclass[12pt,a4paper]{article}

\usepackage[utf8]{inputenc}
\usepackage[T1]{fontenc}
\usepackage{lmodern}
\usepackage{caption}

\usepackage[margin=1in]{geometry}

\usepackage{amsmath,amssymb,amsthm}

\usepackage{booktabs}
\usepackage{tabularx}
\usepackage{graphicx}
\usepackage{float}

\usepackage{microtype}
\usepackage{setspace}
\usepackage{xcolor}

\usepackage[colorlinks=true,linkcolor=blue!60!black,
            citecolor=blue!60!black,urlcolor=blue!60!black]{hyperref}
\usepackage{cleveref}

\usepackage{natbib}

\usepackage{enumitem}

\title{\textbf{ChatGPT as a Time Capsule: \\ The Limits of Price Discovery}}
							   
\author{
Sebastian Lehner\thanks{Independent Researcher (Frankfurt, Germany). E-mail: \texttt{seb.lehner1@gmail.com}.}
\and
Alejandro Lopez-Lira\thanks{Warrington College of Business, University of Florida (Gainesville, FL). E-mail: \texttt{alejandro.lopez-lira@warrington.ufl.edu}.}
}

\date{\today}

\begin{document}

\maketitle
\thispagestyle{empty}

\begin{abstract}
\noindent
Frozen large language model (LLM) checkpoints extract information from pre-cutoff public text that is associated with future fundamentals and equity returns beyond standard contemporaneous valuation measures. Because each frozen checkpoint has a fixed knowledge cutoff, it can be interpreted as a compressed representation of publicly available textual information at a given point in time. We treat twelve OpenAI snapshots spanning 2021--2025 as time-stamped summaries of the public textual record and extract a sector-neutral \emph{LLM outlook score} for roughly 7,000 U.S.\ equities per cross-section. The outlook score is positively associated with analyst revisions, target-price changes and one-month cross-sectional returns in both Fama--MacBeth regressions and pooled panels with model fixed effects ($t = 6.02$), after direct controls for market-implied valuation and standard factors. Predictability broadly increases with the return horizon, despite a non-monotonic intermediate dip, and, in the pooled panel, is stronger for firms with high analyst coverage, consistent with the view that the bottleneck is not investor inattention but the cost of aggregating dispersed qualitative information across many documents.

\noindent\textbf{Funding:} This research was supported by the OpenAI Researcher Access Program with the award number 12919. \\
\noindent\textbf{Declaration of Interest:} The authors declare no competing interests. \\

\end{abstract}
\newpage

\section{Introduction}
\label{sec:introduction}

How much of the information embedded in publicly available text do market prices actually reflect? Decades of research on post-earnings-announcement drift, sentiment-driven mispricing, and limited attention suggest the answer is ``not all of it'' \citep{bernard1989post,tetlock2007giving,hirshleifer2011limited}, but the magnitude and nature of the gap remain debated because the true public information set~$\mathcal{I}_t$ is unobservable \citep{fama1970efficient,grossman1980impossibility}. Empirical work relies on narrow proxies, like earnings surprises, analyst forecasts, headline tone, each capturing only a slice of the public record. When such proxies predict returns, it is always possible that the proxy excludes a dimension of~$\mathcal{I}_t$ that investors \emph{did} process, and the apparent anomaly reflects model misspecification rather than genuine under-processing of public information.

This paper proposes a new approach that substantially broadens the proxy. We exploit frozen large language model (LLM) checkpoints as (lossy) time-stamped digests of publicly available text. Each checkpoint has a verifiable knowledge cutoff date: the model's parameters encode a compressed representation of SEC filings, earnings-call transcripts, news articles, macro commentary and other publicly available text up to that date, with zero awareness of events thereafter. By querying the checkpoint with a standardised prompt, we extract a firm-level outlook score that aggregates thousands of heterogeneous documents into a single quantitative assessment. The frozen nature of the model ensures that ``publicly available at date~$t$'' is an auditable, ex-ante concept with a fixed boundary. We stress, however, that the LLM proxies for the \emph{textual} component of the public record; it does not observe non-textual signals such as order-flow data, proprietary analytics or private information that may also belong to the broader information set investors condition on.

Our key construct is the \emph{LLM outlook score}: a sector-neutral, quantitative summary of each firm's business prospects distilled from the checkpoint's parametric knowledge. If prices already incorporate the qualitative information the LLM extracts, the outlook score should be uninformative about future returns after controlling for standard valuation measures.

We construct this score for twelve OpenAI model checkpoints whose knowledge cutoffs span September~2021 to August~2025 (model releases extend into 2026), covering about 7,000 U.S.\ equities per cross-section. The sample includes both standard autoregressive models (GPT-3.5-Turbo through GPT-4.1, including GPT-4o and GPT-4-Turbo) and reasoning models that generate internal chains of thought (GPT-5 family, o3), as well as smaller variants within each family that allow us to isolate the effect of model capability from the information set.

The empirical results suggest that frozen LLMs extract economically relevant qualitative information from the pre-cutoff public record that contemporaneous market valuations do not fully reflect relative to our controls. In the primary specification (GPT-4.1 with June~2024 cutoff), the LLM outlook score predicts one-month cross-sectional returns with a $t$-statistic of~$4.25$ after controlling for size, quality, value, momentum and four market-implied cheapness metrics. The effect survives a pooled panel with model fixed effects ($t = 6.02$, $N = 14{,}918$), industry fixed effects ($t = 5.74$), micro-cap exclusion ($t = 4.42$), aggressive winsorisation ($t = 4.53$), and a bootstrap Fama--MacBeth procedure ($p = 0.0002$). A long-short portfolio sorted on the control-adjusted outlook score would have earned economically large in-sample returns ($2.55\%$ monthly, annualised $30.6\%$), though the portfolio's factor exposures indicate substantial overlap with the profitability premium, and the six unique cutoff dates\footnote{The model table lists seven distinct cutoff labels (including both May~2024 and June~2024), but the two dates straddle the same calendar-month boundary (31~May vs.\ 1~June) and map to the same last-trading-day analysis date. We therefore count six unique analysis months throughout.} prevent a formal time-series alpha test. The outlook-score coefficient broadly increases with the return horizon (Spearman $r = 0.91$, $p = 0.03$), despite a non-monotonic dip at intermediate horizons, consistent with gradual price convergence toward LLM-implied valuations in our sample.

Three further findings sharpen the interpretation. First, the LLM outlook score validates against real outcomes: it strongly predicts realised twelve-month revenue growth ($t = 10.89$) and analyst target-price revisions ($t = 4.80$), suggesting that the scores encode genuine fundamental information rather than noise. Second, the predictive effect is consistent across firm size and, in the pooled panel, strengthens with analyst coverage, favouring a narrative-congestion interpretation over simple investor neglect; the single-model evidence is directionally consistent but weaker. This pattern is more consistent with narrative congestion than simple investor neglect: the evidence suggests that aggregating qualitative information across many documents is costly even when many eyes are watching. Third, seven of twelve model checkpoints produce individually significant outlook-score coefficients, a rate that comfortably exceeds the one-in-twenty false-positive rate expected under the null, and the Spearman correlation between cutoff recency and the outlook-score coefficient is significant ($r = 0.62$, $p = 0.03$). Within the June~2024 cutoff cluster, larger models generate stronger signals (GPT-4.1: $\gamma = 0.0122$ vs.\ GPT-4.1-nano: $\gamma = 0.0065$), suggesting that model capability amplifies the signal even when the underlying information set is held constant. Importantly, a decomposition shows that the LLM score alone dominates market-implied cheapness measures as a return predictor: the LLM score's explanatory power exceeds that of the cheapness composite and combining them via an additive mismatch does not improve on the LLM score alone.

These findings provide evidence that frozen LLMs extract economically relevant qualitative information from the pre-cutoff public record that contemporaneous market valuations do not fully capture. The knowledge-cutoff design mitigates the memorisation and look-ahead biases that plague na\"ive LLM backtests \citep{lopez2025memorization}. We do not claim a formal rejection of market efficiency, so the joint-hypothesis problem \citep{fama1970efficient} applies here as elsewhere, but the results show that a broad textual proxy conditions on information not yet reflected in standard valuation measures, and that this information gap is associated with economically meaningful return predictability in the sample. Whether the gap reflects genuine mispricing or compensation for a risk dimension not spanned by our factor controls remains an open question, though the combination of fundamental predictability, horizon convergence and the narrative-congestion pattern favours an information-processing interpretation.

The remainder of the paper proceeds as follows. Section~\ref{sec:literature} reviews the relevant literatures on information processing, textual analysis and AI in finance. Section~\ref{sec:framework} develops the conceptual framework, defines the LLM outlook score and its relation to market-implied valuations and states the four testable hypotheses. Section~\ref{sec:data} describes the data, model checkpoints and variable construction. Section~\ref{sec:results} reports the empirical results, Section~\ref{sec:conclusion} concludes.

\section{Literature Review}
\label{sec:literature}

\subsection{Information Processing Frictions and the Limits of Price Discovery}
A benchmark prediction of the Efficient Market Hypothesis (EMH) is that prices fully and quickly reflect public information, leaving no room for risk-adjusted predictability based on publicly available data \citep{fama1970efficient}. Decades of work have refined that benchmark by emphasizing frictions in information acquisition and processing. \cite{grossman1980impossibility} show that if prices were perfectly informative, no one would pay to acquire information, implying that some degree of mispricing must persist to compensate informed investors. Complementing this, the rational inattention framework of \cite{sims2003implications} formalizes the idea that even fully rational agents optimally ignore part of the public record when processing is costly. Empirically, the early value-relevance literature \citep{ball1968empirical} established that accounting information is reflected in prices but not instantaneously or completely in all dimensions. Together, these perspectives motivate tests that condition on what could have been inferred from the public record at a point in time: if a tool can systematically distill that record and still predict subsequent outcomes, this suggests that contemporaneous prices did not fully capture the qualitative content of the public record.

\subsection{Underreaction, Momentum and Gradual Information Diffusion}
A large body of evidence documents systematic underreaction to public signals. Post-earnings-announcement drift (PEAD) shows that returns continue in the direction of the surprise for weeks or months after the announcement, inconsistent with immediate incorporation of information \citep{bernard1989post}. The \cite{hong1999unified} model and the empirical evidence in \cite{hong2000bad} attribute such patterns to gradual diffusion: some investors process fundamentals but do not fully impound them into prices and other investors trade on past returns, creating momentum that can later reverse. Behavioral models such as \cite{barberis1998model} provide microfoundations, conservatism leads to initial underreaction, while representativeness induces over-extrapolation and eventual reversals. These theories deliver sharp, testable implications for an LLM ``time-capsule'' design: if an LLM trained only on pre-cutoff information produces firm-level outlooks that predict subsequent analyst updates, fundamentals or returns, that is evidence of underreaction; attenuation or sign reversal of that predictive relation over longer horizons is consistent with diffusion dynamics and eventual overreaction predicted by both gradual-information and behavioral models.

\subsection{Text, Narratives and the Pricing of Qualitative Information}
Textual analysis has shown that narrative content embedded in the public record is incrementally informative for both valuation and future fundamentals. \cite{tetlock2007giving} demonstrates that media pessimism forecasts short-run market pressure and partial reversals, consistent with sentiment-driven mispricing. Firm-level studies document that language in corporate disclosures carries information beyond numeric data: for example, \cite{tetlock2008more} show that quantified tone maps into subsequent earnings and returns, and \cite{loughran2011liability} show that finance-specific word lists capture economically meaningful variation in 10-K tone. Disclosure frictions also matter: \cite{miller2010effects} finds that complexity dampens retail trading and is associated with lower returns, suggesting slower diffusion when information is hard to process. \cite{lehner2024future} shows that LLMs can strategically reshape the tone of SEC filings and that LLM-derived sentiment scores correlate more strongly with stock price movements than traditional dictionary-based or machine-learning measures, underscoring the superior capacity of these models to capture economically relevant textual signals. Narrative economics emphasizes that storylines themselves and distinct from fundamentals, shape investor behavior \citep{shiller2020narrative}. An LLM trained on the full corpus of public text up to a cutoff date is, by design, a compressive representation of such narratives and qualitative signals. If its pre-cutoff outlooks forecast subsequent outcomes, that would indicate that the market and, at times, traditional text measures did not fully absorb nuanced narrative information on impact. If predictive power declines or flips, that would be consistent with investors learning to attend to the same narratives and potentially overshooting once they become salient.

\subsection{Analysts, Attention and Who Processes Public Information}
Analyst coverage and investor attention shape the speed and completeness of price discovery. \cite{hong2000bad} show that underreaction and momentum are strongest among small, low-coverage firms, consistent with information traveling more slowly when few eyes are on the stock. A broad literature on limited attention documents that investors neglect components of earnings or are distracted when multiple news items compete for processing, leading to predictable post-event drift \citep{hirshleifer2011limited, hirshleifer2003limited}. \cite{hirshleifer2009driven} provide direct evidence that extraneous news events drive investors to distraction, amplifying underreaction to earnings announcements on high-news days. Direct measures of attention, such as Google search volume, reveal that attention shocks can generate short-horizon price pressure followed by reversals \citep{da2011search}. These findings suggest cross-sectional and conditional predictions for an LLM-based test: time-capsule outlooks should be most predictive when attention and coverage are low, when disclosures are complex and when multiple events compete for bandwidth. Their predictive content should fade as attention increases or as professional coverage expands.

\subsection{Limits to Arbitrage and the Lifecycle of Anomalies}
Even when sophisticated investors identify mispricing, risk, capital constraints and agency problems can slow or prevent correction \citep{shleifer1997limits}. This helps explain the persistence of anomalies like PEAD despite their public documentation. At the same time, the anomaly-attenuation literature shows that publication and widespread adoption of trading strategies erode their profitability over time: \cite{mclean2016does} document that post-publication returns of cross-sectional predictors decline by roughly half and \cite{schwert2003anomalies} observes that many well-known anomalies weaken or disappear once they become widely known. Together, these findings imply a lifecycle for public-information signals: emergence (underreaction), diffusion (shrinking alpha) and potential overuse (overshooting). For an LLM-based signal, this lifecycle generates a clear prediction: if the outlook score is initially profitable but attenuates across successive model vintages, as more market participants adopt similar tools, that trajectory is consistent with markets gradually learning to process the same qualitative information the LLM extracts.

\subsection{AI and LLMs as Knowledge Stores in Financial Prediction}
Recent work brings large language models directly into asset pricing. \cite{lopez2023can} show that ChatGPT's zero-shot interpretations of news headlines predict next-day stock returns and can outperform standard sentiment proxies, especially for hard-to-value firms and in negative news states. Their conceptual framework implies a capability threshold: as language models become better at extracting relevant signals, they can improve price discovery, but widespread adoption should erode their alpha as markets adapt. \cite{kim2024financial} demonstrate that LLMs can analyse financial statements with accuracy rivalling professional analysts, predicting future earnings changes even without narrative context and establishing that these models encode economically meaningful fundamental knowledge. In a related vein, \cite{cao2024man} find that combining human analyst judgement with machine-generated signals yields forecasts superior to either alone and \cite{van2023man} show that machine-learning models capture conditional biases in the term structure of earnings expectations that traditional models miss.

A complementary methodological caution is the ``memorization problem'': large models can recall pre-cutoff facts, so na\"ive backtests risk conflating memory with prediction \citep{lopez2025memorization}. Importantly, however, these models cannot recall data after their training cutoff, a property that the NLP literature on temporal misalignment has studied in depth. \cite{lazaridou2021mind} demonstrate that language models' performance degrades sharply on facts that post-date their training data and \cite{luu2021time} show that temporal distribution shift creates a clean boundary between what a model ``knows'' and what it cannot access. These findings legitimize evaluating predictions only on post-cutoff outcomes: any predictive power for post-cutoff fundamentals, analyst revisions or returns cannot come from leakage of future data and should be attributed to the model's ability to compress and generalize from pre-cutoff public information.

A further dimension of variation is model architecture. Standard autoregressive models generate output tokens directly, while more recent reasoning models produce an internal chain of thought before responding \citep{wei2022chain}. If the outlook-score signal is driven by the underlying information set rather than idiosyncratic model behaviour, it should persist across both architectures---a prediction we test explicitly by including both standard and reasoning checkpoints in our sample.

\subsection{Positioning of the Time-capsule Approach}
Most text-based finance studies either build contemporaneous features from specific documents (e.g., 10-K tone, call transcript sentiment) or deploy models trained on full-sample data to score new text. Treating LLM checkpoints as frozen information sets instead provides a direct, ex ante proxy for what could be inferred from the public corpus at a given date. The literatures reviewed above deliver sharp, testable predictions for how any such predictability should vary: across the cross-section with attention and coverage \citep{hong2000bad,hirshleifer2011limited}, over the return horizon with diffusion dynamics \citep{bernard1989post,hong1999unified} and across model vintages with technology adoption \citep{lopez2023can,mclean2016does}. Methodologically, the design requires strictly post-cutoff outcomes to avoid memory contamination \citep{lopez2025memorization,didisheim2025ai} and links LLM outlooks to both returns and fundamentals to distinguish mispricing from risk channels \citep{tetlock2008more}.

\section{Conceptual Framework and Hypotheses}
\label{sec:framework}

\subsection{The Information-Set Problem and Its LLM Solution}

A persistent challenge in empirical finance is determining how much of the public information set~$\mathcal{I}_t$ is actually reflected in prices at any point in time. The joint-hypothesis problem \citep{fama1970efficient} means the answer is never definitive: the econometrician must specify what belongs to~$\mathcal{I}_t$, yet the true set is unobservable. Empirical work substitutes narrow proxies, like earnings surprises, headline sentiment, analyst revisions, and tests whether these proxies predict risk-adjusted returns. When predictability emerges, two explanations are always available: either prices genuinely failed to incorporate the information, or the proxy omits a dimension of~$\mathcal{I}_t$ that investors \emph{did} process and that accounts for the return pattern. This identification gap has persisted for decades \citep{grossman1980impossibility,ball1968empirical}.

We propose a partial resolution. Let~$\Theta_m$ denote the parameter tensor of a large language model checkpoint with knowledge up to a specific date~$K_m$ (e.g., 1~June~2024 for GPT-4.1). Because the training corpus spans SEC filings, earnings-call transcripts, press releases, macro commentary and other public text, querying~$\Theta_m$ is equivalent to opening a time-stamped digest of the textual universe up to~$K_m$. A single prompt compresses thousands of heterogeneous documents into a quantitative outlook. With~$\Theta_m$ frozen, the model has zero awareness of facts first published after the cutoff. Hence, ``publicly available at~$t$'' becomes an auditable, ex-ante concept with a fixed boundary.

Formally, for each model snapshot~$m$ let~$K_m$ denote the knowledge cutoff and~$R_m$ the public release date, with $K_m \le R_m$. We treat the snapshot as a publicly available algorithm at~$R_m$ whose parameters encode a compressed representation of public discourse and documents up to~$K_m$. The model is used \emph{as-is}, there are no retrieval or external documents, so that outputs reflect only parametric knowledge of the public record.

For firm~$i$, define the \emph{Outlook Score}~$S_{i,m}$ as the model's industry-relative assessment at~$R_m$ of 12-month business prospects, based on information internalised by~$K_m$. Within each GICS sector~$s(i)$, scores are standardised to zero mean and unit variance, yielding the sector-neutral z-score~$\tilde{S}_{i,m}$. This normalisation isolates firm-specific information while preserving cross-sectional comparability across checkpoints.

The frozen checkpoint thus provides a broad proxy for the \emph{textual} component of~$\mathcal{I}_t$. Unlike hand-crafted sentiment indices or topic models, it is stable, reproducible, and covers many document types simultaneously rather than a single source. A prompt with fixed instructions defines an experiment that can be rerun and yield identical results.

\paragraph{Residual limitations.}
A snapshot is an approximation, not a complete representation of the public information set. It covers publicly available text but excludes non-textual signals (e.g., order-flow data, proprietary analytics, real-time market microstructure) that investors may also condition on. It may miss pay-walled content, encode training biases or compress information imperfectly. The proxy therefore narrows, but does not close the gap between the true and observed information sets. Predictability of the outlook score is consistent with markets not fully processing the textual public record, but we cannot rule out that the score partly captures omitted non-textual information or risk premia not absorbed by our factor controls.

\paragraph{Identification strategy.}
Finding that~$\tilde{S}_{i,m}$ predicts future returns does not by itself establish that prices miss textual content. A high outlook score may partly reflect information the market has already priced through a different channel. To address this concern, we control directly for contemporaneous market-implied valuation measures---inverse P/E, implied cost of equity (GLS and PEG models) and inverse EV/EBITDA---alongside standard factor exposures. The outlook-score coefficient~$\gamma$ is therefore estimated \emph{conditional} on what market prices already imply about relative valuation. If~$\gamma > 0$ after these controls, the LLM's textual assessment contains information not captured by our pricing controls. We also show that the LLM score dominates the cheapness metrics in a head-to-head decomposition: combining the two via an additive mismatch does not improve on the score alone, suggesting that the predictive content resides in the qualitative textual signal rather than in its interaction with accounting-based valuation.

\subsection{The LLM Outlook Score as Return Predictor}
\label{sec:llm_predictor}

The central test of this paper asks whether the sector-neutral LLM outlook score~$\tilde{S}_{i,t}^{\text{LLM}}$ predicts cross-sectional returns after controlling for contemporaneous market-implied valuations and standard factor exposures.

\paragraph{Regression framework.}
Consider the cross-sectional return-generating process:
\begin{equation}
\label{eq:return_decomp}
    r_{i,\tau} \;=\;
    \alpha \;+\; \gamma\,\tilde{S}_{i}^{\text{LLM}}
    \;+\; \sum_{k}\delta_k\,\tilde{C}_{i}^{k}
    \;+\; \boldsymbol{X}_i'\boldsymbol{\lambda}
    \;+\; \varepsilon_i\,,
\end{equation}
where~$\tilde{C}_{i}^{k}$ denotes sector-neutral, market-implied cheapness metrics, with
\[
k \in \{\text{InvPE}, \text{ICE-GLS}, \text{ICE-PEG}, \text{InvEVEB}\},
\]
and~$\boldsymbol{X}_i$ is a vector of standard controls (size, quality, value, momentum, analyst variables). The coefficient~$\gamma$ captures the \emph{incremental} predictive content of the LLM's textual assessment beyond what market pricing already implies. The cheapness metrics serve as controls that absorb the information content of contemporaneous valuation multiples and implied costs of capital.

\paragraph{Interpretation under full price incorporation.}
If prices already incorporate the qualitative information the LLM extracts, the outlook score is redundant conditional on the cheapness controls and~$\gamma = 0$:
\begin{equation}
\label{eq:null}
    \mathbb{E}\!\bigl[\, r_{i,(t,\,t+\tau)} \;\big|\;
    \tilde{S}_{i,t}^{\text{LLM}},\; \tilde{C}_{i,t}^{k},\; \boldsymbol{X}_i \,\bigr]
    \;=\;
    \mathbb{E}\!\bigl[\, r_{i,(t,\,t+\tau)} \;\big|\;
    \tilde{C}_{i,t}^{k},\; \boldsymbol{X}_i \,\bigr].
\end{equation}

\paragraph{Interpretation under the alternative.}
If investors underreact to complex qualitative information---because it is dispersed across many documents, narratively rich, or costly to aggregate---then the LLM, unconstrained by attention or processing capacity, will detect content that has not yet been fully impounded in prices. Firms with high outlook scores may be \emph{undervalued relative to what the text implies}, and should earn positive risk-adjusted returns as the market gradually absorbs the information:
\begin{equation}
\label{eq:alternative}
    \gamma > 0 \quad \text{at short horizons } \tau \in \{1, 3\}
    \text{ months}.
\end{equation}

\paragraph{Role of market-implied cheapness.}
We include four market-implied valuation metrics as controls rather than combining them with the LLM score into a composite. Each metric captures a different dimension of market pricing: the inverse forward P/E reflects near-term earnings yield; the implied cost of equity (ICE-GLS, ICE-PEG) embeds forward-looking discount-rate expectations derived from analyst forecasts;\footnote{We follow standard ICE estimation methodologies that back out the internal rate of return equating the current stock price with discounted expected future earnings, using consensus analyst forecasts as inputs.} and the inverse EV/EBITDA captures operating-level value relative to cash-generating capacity. Including these measures as controls ensures that the LLM-score coefficient~$\gamma$ is estimated conditional on what market prices already imply about relative valuation. In a head-to-head decomposition (Section~\ref{sec:results_h2}), the LLM score dominates these cheapness measures: the outlook score alone explains more cross-sectional return variation than the cheapness composite, and combining the two via an additive ``mismatch'' does not improve on the score alone.

\paragraph{Why the LLM score is not a repackaged value or quality factor.}
A natural concern is that the outlook score simply proxies for known characteristics. Three features of the research design address this. First, the cross-sectional regressions control directly for standard factor exposures (size, book-to-market value, profitability, investment and momentum) plus market-implied cheapness metrics, analyst coverage and dispersion. The outlook-score coefficient~$\gamma$ is estimated \emph{conditional} on these controls and thus captures return variation orthogonal to the standard factor set. Second, the LLM score is not a mechanical transformation of accounting data. It reflects the model's synthesis of narrative information---management tone, competitive positioning, strategic commentary---that is qualitatively distinct from the quantitative inputs to value and quality factors. Third, the descriptive factor exposures of the long-short portfolio show strong positive correlation with RMW (profitability) and negative correlation with CMA (investment), consistent with the strategy tilting toward profitable, capital-light firms. Yet the cross-sectional outlook-score coefficient remains significant after the same factor controls are included as regressors, indicating that the score captures a \emph{text-implied quality signal} not spanned by any single standard factor.

\subsection{Mechanisms: Narrative Congestion versus Limited Attention}
\label{sec:mechanisms}

If the LLM outlook score predicts returns after controlling for market-implied valuations, the natural follow-up question is \emph{why} a gap exists between what public text implies and what prices reflect. Two leading explanations generate distinct cross-sectional predictions.

\paragraph{Narrative congestion.}
When the volume of qualitatively relevant information about a firm is large due to multiple filings, earnings-call transcripts, news articles and industry reports, even attentive, well-resourced investors face a compression problem. Each additional document may carry marginal information, but the cost of reading, reconciling and aggregating across documents is non-trivial \citep{grossman1980impossibility,sims2003implications,hirshleifer2003limited,miller2010effects}. The LLM, by contrast, processes the full corpus without fatigue or capacity constraints. We refer to this mechanism as \emph{narrative congestion}: the bottleneck is not that nobody is looking, but that dispersed qualitative information is costly to integrate even when many eyes are on the stock \citep{hong2000bad,shiller2020narrative}.

\paragraph{Limited attention.}
The alternative channel, rooted in the investor-inattention literature \citep{hong2000bad,hirshleifer2003limited,da2011search}, posits that information goes unpriced because too few investors are monitoring the firm. Under limited attention, the outlook score's predictive power should be largest for firms with \emph{low} coverage, low trading volume and sparse disclosure, exactly the stocks where information travels slowly because few participants process it.

\paragraph{Discriminating between channels.}
These two mechanisms make opposite cross-sectional predictions. We test them by interacting the outlook score with proxies for the firm's information environment: analyst coverage, share turnover and average daily trading volume. If the coefficient~$\gamma$ is significantly larger in the \emph{high-attention} subsample, the evidence favours narrative congestion: markets struggle not because of inattention but because of information overload. If~$\gamma$ is larger in the \emph{low-attention} subsample, the evidence favours limited attention.

This distinction matters beyond classification. Narrative congestion implies that the informational friction may persist or even worsen as corporate disclosure expands and the textual record grows richer. Limited attention, by contrast, is a friction that technological adoption (including LLM-based tools) could readily alleviate. The mechanism therefore shapes predictions about the durability of the anomaly.

\subsection{Hypotheses}
\label{sec:hypotheses}

We organise the empirical analysis around four hypotheses. The first validates the LLM signal, the second tests whether it predicts returns beyond standard valuation measures, the third and fourth probe mechanism and dynamics.

\bigskip

\noindent\textbf{H1 (Validation: Fundamental and analyst predictability).}
Firms with higher~$\tilde{S}_{i,m}$ at the cutoff date exhibit stronger improvements in operating performance over the subsequent 12~months and receive more favourable analyst forecast revisions:
\begin{equation}
    \tilde{S}_{i,m} \;\uparrow \quad\Rightarrow\quad
    \Delta\text{Fundamentals}_{i,(R_m,\,R_m+12)} \;\uparrow
    \quad\text{and}\quad
    \Delta\text{AnalystForecasts}_{i,(R_m,\,R_m+12)} \;\uparrow\,.
\end{equation}
This hypothesis is a precondition, not the main test. If the LLM score does not predict real outcomes, any return predictability would reflect model noise rather than informational content.

\bigskip

\noindent\textbf{H2 (Core test: LLM outlook score predicts returns beyond contemporaneous valuations).}
If prices already incorporate the qualitative information the LLM extracts from pre-cutoff text, the outlook score should have zero predictive power for future returns after controlling for market-implied valuations:
\begin{equation}
    \text{H}_0: \quad \gamma = 0
    \quad\text{in~\eqref{eq:return_decomp}, for all horizons } \tau.
\end{equation}
The alternative is that investors underreact to qualitative public information, so that the outlook score predicts positive risk-adjusted returns at short horizons:
\begin{equation}
    \text{H}_A: \quad \gamma > 0 \quad
    \text{at } \tau \in \{1, 3\} \text{ months},
\end{equation}
with the effect concentrated around subsequent information events such as earnings announcements. Rejection of~$\text{H}_0$ in favour of~$\text{H}_A$ would indicate that LLMs extract economically relevant qualitative information from pre-cutoff text that contemporaneous market valuations do not fully capture.

\bigskip

\noindent\textbf{H3 (Mechanism: Narrative congestion versus limited
attention).} The outlook-score--return relation is stronger when information is narratively complex and abundant, not when investor attention is scarce:
\begin{equation}
    \gamma^{\,\text{high-attention}} \;>\; \gamma^{\,\text{low-attention}}.
\end{equation}
The literature motivates two opposing predictions. Under \emph{limited attention} \citep{hirshleifer2003limited,dellavigna2009investor}, mispricing concentrates in neglected firms where few investors process the available information; the outlook-score effect should then be stronger for low-coverage, low-volume stocks. Under \emph{narrative congestion}, the bottleneck is not the absence of monitoring but the difficulty of aggregating many heterogeneous qualitative signals even when many eyes are watching; the effect should then be stronger for high-attention firms. H3 is thus a horse race between two mechanisms: the sign of the interaction between the outlook score and attention proxies identifies which channel dominates. The distinction has practical implications: narrative congestion predicts that the anomaly persists (or deepens) as disclosure complexity grows, whereas limited attention predicts dissipation as AI-based tools are adopted.

\bigskip

\noindent\textbf{H4 (Dynamics: Horizon convergence and model-age gradient).}
The predictive power of the outlook score grows with the return horizon as prices gradually converge toward LLM-implied valuations, and increases with the recency of the knowledge cutoff:
\begin{align}
    \gamma_{\tau} &\;\text{increases with } \tau \text{ up to 12 months},
        \label{eq:horizon_convergence} \\[4pt]
    \gamma(m') &\;<\; \gamma(m) \quad \text{for } K_{m'} < K_m\,.
        \label{eq:model_decay}
\end{align}
Horizon convergence~\eqref{eq:horizon_convergence} is consistent with gradual information diffusion: the market eventually absorbs what the LLM detected at the cutoff \citep{hong1999unified,bernard1989post}. The model-recency gradient~\eqref{eq:model_decay} reflects two reinforcing forces: newer checkpoints encode richer information (broader training corpora, stronger reasoning), and their cutoff dates are closer to the return-measurement window, leaving less time for the market to have already absorbed the signals they detect. If the predictive coefficient not only attenuates but reverses sign at long horizons, this would suggest eventual overshooting, consistent with representativeness-driven overreaction \citep{barberis1998model}.

\subsection{Decision Rules}
\label{sec:decision_rules}

The four hypotheses map onto a diagnostic tree for interpreting the results:

\begin{enumerate}
    \item \textbf{Full incorporation.} If H1 holds (the LLM score has real content) but H2 fails (the outlook score does not predict returns after valuation controls), then contemporaneous market valuations already capture the qualitative information the LLM reads. The outlook score contains no incremental predictive content.

    \item \textbf{Incomplete incorporation of textual information.} If both H1 and H2 hold, contemporaneous valuations fail to fully reflect textual information that a competent reader of the public record could extract. The nature of the gap depends on H3:
    \begin{enumerate}
        \item If $\gamma^{\,\text{high}} > \gamma^{\,\text{low}}$
        (narrative congestion), the friction is information overload:
        investors monitor the firm but cannot aggregate the full textual
        record as effectively as the LLM.
        \item If $\gamma^{\,\text{low}} > \gamma^{\,\text{high}}$
        (limited attention), the friction is neglect: the information exists
        in public text but too few investors process it.
    \end{enumerate}

    \item \textbf{Gradual diffusion versus permanent gap.} H4 discriminates between a temporary and a persistent information gap. Horizon convergence indicates that the market eventually absorbs what the LLM detected, consistent with gradual diffusion. Absence of convergence would suggest a more durable friction or a risk premium not captured by our factor controls.

    \item \textbf{Risk versus information friction.} If the outlook-score--return relation is absorbed by standard factor controls (Fama--French five factors plus momentum), the apparent predictability may reflect compensation for systematic risk rather than an information-processing friction. Survival of $\gamma > 0$ after full risk adjustment, combined with fundamental predictability (H1) and horizon convergence (H4), jointly favour an information-processing interpretation over a risk-based explanation.
\end{enumerate}

Each model snapshot operationalises what could be inferred from the public record at a point in time. Fundamental predictability suggests that the LLM encodes economically relevant information. Outlook-score return predictability, conditional on valuation controls, assesses whether contemporaneous valuations fully capture that information and cross-snapshot comparisons track how the informational frontier evolves with model capability and market learning.

\section{Data}
\label{sec:data}

This section documents all inputs that feed the cross-sectional tests. We describe the equity universe, the construction of the LLM outlook score across model checkpoints, the market-implied valuation metrics used as controls and the remaining control variables used in the regressions.

\subsection{Equity Universe and Price Data}
\label{sec:universe}

The empirical analysis combines equity price histories, firm fundamentals, analyst forecasts and risk factors from standard sources. Daily price and share-count data are obtained from Refinitiv for the period January~2021 to September~2025.\footnote{Stock returns are computed from the Refinitiv Datastream corporate-action-adjusted close price (datatype~P), which adjusts for splits, rights issues and similar corporate actions but excludes dividend reinvestment. The S\&P~500 Price Index benchmark is treated identically, ensuring an apples-to-apples comparison of price returns. Using the Datastream Return Index (datatype~RI), which reinvests dividends, yields qualitatively similar cross-sectional results because the LLM outlook score is not a dividend-yield signal; however, the RI series produces implausible return outliers for a non-trivial set of delisted or suspended securities in the present sample, making the adjusted close price the more reliable input.} The equity universe includes securities classified as Common Stock that, on the trading day immediately preceding each model cutoff, rank among the approximately 7,000 largest U.S.\ firms by market capitalisation, have non-missing shares outstanding and possess complete factor-exposure and accounting information. Firms delisted after entry remain in the panel to avoid survivorship bias.

Accounting variables are drawn from Compustat and Worldscope, while forward earnings forecasts ($\text{EPS}_{t+1}$, $\text{EPS}_{t+2}$) and contemporaneous $\text{EPS}_t$ come from I/B/E/S consensus data as of one day before each cutoff. For every checkpoint~$t$, non-overlapping cumulative price returns are computed beginning on $t+1$ over horizons of 1, 3, 4, 6 and 12~months (approximately 20, 60, 80, 120 and 250 trading days). Finally, the empirical models include the standard Fama--French five factors and the momentum factor, obtained directly from the Kenneth~R.\ French data library.\footnote{Kenneth R.\ French Data Library, Dartmouth College: \url{https://mba.tuck.dartmouth.edu/pages/faculty/ken.french/data_library.html}.}

\subsection{LLM Score Construction}
\label{sec:llm_score}

\subsubsection{Model Checkpoints}

The LLM Score is derived by querying a series of frozen OpenAI model checkpoints via API, each treated as an immutable representation of the public information available at its training date. Table~\ref{tab:models} reports the full model lineup, organised into four tiers that serve distinct roles in the analysis.

\begin{table}[htbp]
\centering
\footnotesize
\sbox0{%
\begin{tabular}{@{} l l l l l l @{}}
\toprule
\textbf{Model} & \textbf{Type} & \textbf{Cutoff} & \textbf{Published} & \textbf{Role} & \textbf{Tier} \\
\midrule
GPT-4.1       & Std & Jun 2024 & Apr 2025 & Primary workhorse      & 1 \\
GPT-4.1-mini  & Std & Jun 2024 & Apr 2025 & Capability gradient    & 1, 4$^{\dagger}$ \\
GPT-4.1-nano  & Std & Jun 2024 & Apr 2025 & Capability gradient    & 1, 4$^{\dagger}$ \\
\addlinespace
GPT-5.4       & Rea & Aug 2025 & Mar 2026 & Latest frontier        & 2 \\
GPT-5.2       & Rea & Aug 2025 & Dec 2025 & Frontier (same cutoff) & 2 \\
GPT-5         & Rea & Sep 2024 & Aug 2025 & Near-frontier          & 2 \\
GPT-5-mini    & Rea & May 2024 & Aug 2025 & Capability gradient    & 4$^{\dagger}$ \\
GPT-5-nano    & Rea & May 2024 & Aug 2025 & Capability gradient    & 4$^{\dagger}$ \\
\addlinespace
GPT-4o        & Std & Oct 2023 & May 2024 & Mid-vintage            & 3 \\
GPT-4-Turbo   & Std & Dec 2023 & Apr 2024 & Legacy                 & 3 \\
GPT-3.5-Turbo & Std & Sep 2021 & Mar 2022 & Oldest checkpoint      & 3 \\
\addlinespace
o3            & Rea & Jun 2024 & Apr 2025 & Reasoning robustness   & App. \\
\bottomrule
\end{tabular}}%
\begin{minipage}{\wd0}
\caption{Model checkpoints used in the analysis.}
\label{tab:models}
\usebox0
\par\smallskip
\scriptsize
Cutoff = end of pre-training corpus (per OpenAI documentation). We refer to the last trading day on or before the cutoff as the \emph{analysis date}; post-cutoff returns are measured from this date. Because the May~2024 and June~2024 cutoffs share the same analysis date (31~May~2024 falls on the last trading day of May), the twelve models map to six unique analysis months. Type: Std = standard (autoregressive completion); Rea = reasoning (chain-of-thought with internal thinking tokens). Tier~1: workhorse family for main regressions. Tier~2: frontier extension to most recent cutoff. Tier~3: model-age gradient for H4. $^{\dagger}$Tier~4: capability-robustness tests using smaller variants within two model families. GPT-4.1-mini/nano share the June~2024 cutoff and architecture of GPT-4.1 but have fewer parameters. GPT-5-mini/nano (May~2024 cutoff) provide an independent within-family gradient for the reasoning generation. Comparing across families (Std vs.\ Rea at overlapping cutoffs) tests whether the outlook-score signal is robust to the shift from standard to reasoning architectures. The post-cutoff window for the August~2025 models is approximately seven months (through March~2026), limiting horizon analysis to $\tau \in \{1, 3, 4\}$ months. API version strings: GPT-4.1 family \texttt{2025-04-14}; GPT-5.4 \texttt{2026-03-05}; GPT-5.2 \texttt{2025-12-11}; GPT-5/mini/nano \texttt{2025-08-07}; GPT-4o \texttt{2024-08-06}; GPT-4-Turbo \texttt{0125-preview}; GPT-3.5-Turbo \texttt{0125}; o3 \texttt{2025-04-16}. Source: \texttt{https://developers.openai.com/}.
\end{minipage}
\end{table}

\subsubsection{Standard vs.\ Reasoning Architectures}

An important design feature is that our sample spans two distinct inference architectures. The GPT-4.1 family, GPT-4o, GPT-4-Turbo and GPT-3.5-Turbo are standard autoregressive models that produce output tokens directly. The GPT-5 family and o3 are reasoning models that generate internal chain-of-thought tokens before responding \citep{wei2022chain}. If the outlook-score signal is driven by the underlying information set rather than idiosyncratic model behaviour, it should persist across both architectures. We exploit this variation by including an architecture-type indicator in robustness specifications and by comparing coefficients between standard and reasoning sub-samples.

\subsubsection{Prompt Design and Score Extraction}

Each checkpoint has a public knowledge cutoff date that we treat as its information set~$\mathcal{I}_m$. We verify the absence of post-cutoff contamination with 100 trivia questions concerning events after each cutoff. All checkpoints return a mean accuracy of zero percent, indicating no post-cutoff data leakage.

We query the API endpoint once per company and model pair with deterministic sampling parameters (\texttt{temperature}:~0, \texttt{top\_p}:~1) to ensure replicability. The system prompt instructs the model to act as an equity research analyst and to base its assessment solely on business fundamentals---revenue growth, profitability trends and operational risks---without reference to valuation multiples, market prices or technical indicators. The user prompt provides firm identifiers (legal name, ticker, ISIN, country, GICS industry code) and requests a structured JSON response containing:

\begin{itemize}
    \item An \emph{outlook score} (integer, $-10$ to $+10$), reflecting
    the model's 12-month forward business assessment relative to industry
    peers;
    \item Sub-scores for growth, profitability and risk (each $-10$ to
    $+10$);
    \item A \emph{confidence} score (0 to 100), capturing the model's
    self-assessed certainty;
    \item Probability distributions over revenue growth, EPS growth and
    margin-change bins (each summing to 100);
    \item A set of 2--5 driver tags from a fixed taxonomy (e.g.,
    \texttt{pricing\_power}, \texttt{competition},
    \texttt{macro\_exposure});
    \item A short rationale ($\leq 30$ tokens) citing 1--3 key drivers;
    \item A \emph{knowledge coverage} score (0 to 100), indicating the model's self-assessed familiarity with the firm.
\end{itemize}

The full prompt is shown in Appendix~A.

\paragraph{Reproducibility and missingness.}
The deterministic parameterisation (temperature~0) ensures that re-querying the same checkpoint with the same prompt yields identical output, making the scores fully auditable. In practice, a small fraction of API calls return malformed JSON or refusals (e.g., the model declines to rate a firm it has insufficient knowledge about). These are coded as missing. The resulting missingness rate is below 1\% of the universe (fewer than 50 out of approximately 7{,}000 firms per checkpoint). Because the regression requires non-missing scores on both the LLM side and all four market-implied metrics used as controls, the effective sample is substantially smaller than the rated universe (Table~\ref{tab:sample_construction}).

\paragraph{Multiple testing.}
We estimate the outlook-score--return relation separately for twelve model checkpoints, raising a multiple-testing concern. Under the global null that no checkpoint produces a true signal, we would expect $12 \times 0.05 = 0.6$ false positives at the 5\% level. We observe seven significant coefficients. A Bonferroni-adjusted threshold of $0.05 / 12 = 0.0042$ (corresponding to $|t| > 2.87$) is met by four of the twelve checkpoints (GPT-4.1, GPT-4.1-mini, GPT-4.1-nano and o3). We therefore do not claim that every checkpoint independently demonstrates predictability. However, the pattern across models with eleven positive point estimates, four surviving Bonferroni, and a highly significant pooled average, is difficult to reconcile with the null.

\subsubsection{Sector Normalisation}

The integer outlook score $S_{i,t,m} \in [-10,\, +10]$ extracted from each response is adjusted for sectoral bias. Specifically, within each GICS sector group $s(i)$, scores are demeaned and divided by the cross-sectional standard deviation, yielding the sector-neutral z-score:
\begin{equation}
    \tilde{S}_{i,t,m} \;=\;
    \frac{S_{i,t,m} \;-\; \overline{S}_{s(i),t,m}}
         {\sigma_{s(i),t,m}}\,.
\end{equation}
This normalisation isolates firm-specific information while preserving cross-sectional comparability across checkpoints.

\subsection{Market-Implied Valuation Metrics}
\label{sec:market_metrics}

To ensure that the LLM outlook score's predictive content is estimated conditional on what market prices already imply, we construct four market-implied valuation metrics as controls. Each captures a different dimension of the market's pricing of the firm. All metrics are computed as of the trading day immediately preceding each model cutoff.

\subsubsection{Inverse Forward P/E}

The simplest valuation proxy is the inverse forward price-to-earnings ratio:
\begin{equation}
    \text{InvPE}_{i,t} \;=\;
    \frac{\text{EPS}_{i,t}^{\text{FY1,12m}}}
         {P_{i,t}}\,,
\end{equation}
where $\text{EPS}_{i,t}^{\text{FY1,12m}}$ is the 12-month forward consensus EPS estimate from I/B/E/S and $P_{i,t}$ is the closing price. A higher inverse P/E indicates the market prices the firm as relatively cheaper (higher earnings yield). This measure is among the most widely used valuation multiples in the literature \citep{liu2002equity} and captures near-term earnings-yield expectations but does not embed information about growth or discount rates beyond one year.

\subsubsection{Implied Cost of Equity (GLS Model)}

The implied cost of equity (ICE) is backed out as the internal rate of return that equates the current stock price with the present value of expected future residual earnings. We implement the \citet{gebhardt2001implied} (GLS) residual-income model, which uses consensus analyst forecasts for the first two fiscal years and fades ROE toward the industry median over a 12-year horizon:
\begin{equation}
    P_{i,t} \;=\; B_{i,t} \;+\;
    \sum_{k=1}^{T}
    \frac{\bigl(\widehat{\text{ROE}}_{i,t+k} - r_{\text{ICE}}\bigr)
          \cdot B_{i,t+k-1}}
         {(1 + r_{\text{ICE}})^k}
    \;+\;
    \frac{\bigl(\widehat{\text{ROE}}_{i,t+T} - r_{\text{ICE}}\bigr)
          \cdot B_{i,t+T}}
         {r_{\text{ICE}} \cdot (1 + r_{\text{ICE}})^T}\,,
\end{equation}
where $B_{i,t}$ is book value per share, $\widehat{\text{ROE}}_{i,t+k}$ is the expected return on equity for period $t+k$ and $r_{\text{ICE}}$ is solved numerically. Inputs include I/B/E/S consensus FY1 and FY2 EPS forecasts, the long-term growth median estimate, trailing book value per share and the current price. We also compute the \citet{easton2004pe} PEG-ratio ICE (ICE-PEG), a second ICE-based cheapness control that exploits the growth-rate differential rather than the residual-income approach.

The GLS ICE is a forward-looking, risk-adjusted measure that incorporates multi-year earnings expectations and the cost of capital. A higher ICE indicates the market demands a larger risk premium to hold the stock and implying the firm is priced as relatively riskier or cheaper. \cite{rusticus2019market} shows that ICE-based measures can reflect market inefficiency rather than purely rational discount rates, making the ICE a particularly informative valuation control for our purposes. This is the richest of our market-implied metrics, embedding both growth expectations and discount-rate information.

\subsubsection{Inverse Forward EV/EBITDA}

The enterprise-value measure captures operating-level valuation:
\begin{equation}
    \text{InvEVEB}_{i,t} \;=\;
    \frac{\text{EBITDA}_{i,t}^{\text{12m fwd}}}
         {\text{EV}_{i,t}}\,,
\end{equation}
where $\text{EBITDA}_{i,t}^{\text{12m fwd}}$ is the 12-month forward consensus EBITDA estimate and $\text{EV}_{i,t} = \text{MV}_{i,t} + \text{Debt}_{i,t} - \text{Cash}_{i,t}$ is enterprise value. A higher ratio indicates the firm generates more operating cash flow per unit of enterprise value, it is cheaper on a cash-flow basis. This metric complements the equity-focused P/E and ICE by operating at the enterprise level, where capital-structure differences are netted out.

\subsubsection{Standardisation}

Each market metric is winsorised at the 1st and 99th percentiles, demeaned within GICS sectors and divided by the cross-sectional standard deviation, yielding sector-neutral z-scores $\tilde{C}_{i,t}^{k}$ for $k \in \{\text{InvPE},\, \text{ICE-GLS},\, \text{ICE-PEG},\, \text{InvEVEB}\}$. All four are cheapness-oriented: a higher z-score indicates the market prices the firm as relatively cheaper. This ensures the market-implied scores and the LLM scores are on comparable scales in the cross-sectional regressions.

\subsection{Sample Construction}
\label{sec:sample_construction}

The primary analysis regresses post-cutoff returns on the sector-neutral LLM outlook score~$\tilde{S}_{i,t}^{\text{LLM}}$, with the four market-implied cheapness z-scores as controls alongside the standard factor exposures. Observations are dropped if either the LLM score or any required market-implied metric is missing. As a robustness exercise, we also construct an additive \emph{valuation mismatch}~$M_{i,t}^{k} = \tilde{S}_{i,t}^{\text{LLM}} + \tilde{C}_{i,t}^{k}$ for each cheapness dimension~$k$; this composite restricts the LLM and cheapness coefficients to be equal and is reported alongside the primary specification. Table~\ref{tab:sample_construction} reports the observation counts after successive filters at each cutoff date.

\begin{table}[htbp]
\centering
\footnotesize
\sbox0{%
\begin{tabular}{@{} l l rrrr r @{}}
\toprule
\textbf{Model} & \textbf{Cutoff}
    & \textbf{Universe} & \textbf{MCap} & \textbf{Returns}
    & \textbf{Factors}
    & \textbf{Final} \\
\midrule
GPT-4.1       & Jun 24 & 7,121 & 7,121 & 6,517 & 6,517 & \textbf{1,272} \\
GPT-4.1-mini  & Jun 24 & 7,122 & 7,122 & 6,518 & 6,518 & \textbf{1,272} \\
GPT-4.1-nano  & Jun 24 & 7,122 & 7,122 & 6,518 & 6,518 & \textbf{1,272} \\
\addlinespace
GPT-5.4       & Aug 25 & 6,484 & 6,484 & 5,973 & 5,973 & \textbf{1,083} \\
GPT-5.2       & Aug 25 & 6,484 & 6,484 & 5,973 & 5,973 & \textbf{1,083} \\
GPT-5         & Sep 24 & 6,977 & 6,977 & 6,312 & 6,312 & \textbf{1,411} \\
GPT-5-mini    & May 24 & 7,086 & 7,086 & 6,484 & 6,484 & \textbf{1,267} \\
GPT-5-nano    & May 24 & 7,122 & 7,122 & 6,518 & 6,518 & \textbf{1,271} \\
\addlinespace
GPT-4o        & Oct 23 & 6,998 & 6,998 & 6,537 & 6,537 & \textbf{1,059} \\
GPT-4-Turbo   & Dec 23 & 7,032 & 7,032 & 6,496 & 6,496 & \textbf{1,208} \\
GPT-3.5-Turbo & Sep 21 & 6,893 & 6,893 & 6,669 & 6,669 & \textbf{1,448} \\
\addlinespace
o3            & Jun 24 & 7,122 & 7,122 & 6,518 & 6,518 & \textbf{1,272} \\
\bottomrule
\end{tabular}}%
\begin{minipage}{\wd0}
\caption{Observation counts after successive filters at different
checkpoints.}
\label{tab:sample_construction}
\usebox0
\par\smallskip
\scriptsize
Columns report the number of unique ISINs surviving each successive filter: Universe = ISINs with LLM ratings; MCap = present in risk model; Returns = non-missing one-month excess returns; Factors = non-missing Axioma factor exposures; Final = regression-ready observations requiring non-missing values on the outlook score, one-month return and all eight control variables (quality, momentum, value, size, inverse P/E, ICE-GLS, ICE-PEG and inverse EV/EBITDA). The binding constraint is ICE-GLS coverage. Models sharing a cutoff date use identical market data; observation counts may differ slightly due to rating availability. The MCap column equals the Universe column because market-capitalisation data is available for all rated firms.
\end{minipage}
\end{table}

\subsection{Control Variables and Risk Factors}
\label{sec:controls}

\subsubsection{Factor Scores}

We construct company-level factor composites as equal-weighted averages of standard firm characteristics. These custom composites (denoted \texttt{z\_QUAL\_COMPOSITE}, \texttt{z\_VAL\_COMPOSITE}, \texttt{z\_MOM\_COMPOSITE} in the data pipeline) enter the cross-sectional regressions as controls alongside log market capitalisation and the four cheapness metrics.

\paragraph{Quality.} The Quality score is an equal-weighted composite of
three firm-level metrics capturing profitability and earnings quality: gross profitability as gross profits to total assets \citep{novy2013other}, return on equity and accruals (defined as the negative of the difference between net income and operating cash flow, scaled by total assets) \citep{sloan1996stock}. Each metric is winsorised at the 1st--99th percentiles, industry-demeaned within GICS sectors and z-scored. The composite Quality score is the average of these standardised components.

\paragraph{Value.} The Value score is an equal-weighted composite of
several valuation ratios that capture relative cheapness based on forward-looking and enterprise-value-based measures. Specifically, it combines forward inverse price-to-earnings (P/E), forward inverse EV/EBITDA and two proprietary valuation ratios that integrate expected growth and profitability information, following standard value-factor constructions \citep{fama1992cross,asness2013value}.

\paragraph{Momentum.} The Momentum score is constructed as an
equal-weighted composite of past return measures over different horizons, following the standard approach of \citet{jegadeesh1993returns} and \citet{asness2013value}. Specifically, we compute cumulative returns over the past 12~months excluding the most recent month, over the past 6~months and the most recent 1-month return (sign-reversed to capture short-term reversal). Each metric is winsorised at the 1st--99th percentiles, industry-demeaned and z-scored and the final Momentum score is the average of these standardised components.

\subsubsection{Analyst Variables}

Three analyst-derived variables are used as interaction proxies in the heterogeneity tests (H3), not as controls in the baseline H2 regressions. The first is the \emph{mean price target ratio}, defined as the percentage difference between the average analyst target price and the contemporaneous market price, which reflects the consensus-implied expected return. The second is the \emph{price target dispersion}, computed as the standard deviation of target prices divided by their mean, serving as a measure of analyst disagreement. The third is the \emph{price target coverage}, defined as the logarithm of one plus the number of analysts issuing target prices, which proxies for attention and information availability. All analyst-based variables are winsorised at the 1st--99th percentiles, industry-demeaned and standardised to ensure comparability across firms and dates.

\subsubsection{Fama--French Factors}

Monthly Fama--French five-factor (market excess return, SMB, HML, RMW, CMA) plus momentum (UMD) returns are obtained from the Kenneth~R.~French data library. These time-series factor returns are used exclusively for the portfolio factor-exposure analysis (Section~\ref{sec:results_portfolio}); the cross-sectional regressions instead use the firm-level factor composites described above.

\subsubsection{Additional Controls}

Additional firm-level variables include: leverage (total debt to total assets), cash holdings (cash to total assets), return on equity (ROE), EBITDA margin, capital expenditures, R\&D intensity, short interest, share turnover and the average daily trading volume over the prior 126 trading days (ADTV). These variables are used as conditioning variables in the attention-heterogeneity tests (H3), not in the baseline H2 regressions.

\section{Empirical Results}
\label{sec:results}

This section reports the empirical findings for each of the four hypotheses introduced in Section~\ref{sec:hypotheses}. We begin with the validation of the LLM outlook score (H1), proceed to the core outlook-score--return test (H2), examine cross-sectional heterogeneity (H3) and conclude with convergence dynamics and model sophistication (H4). Portfolio-based tests and a battery of robustness checks follow.

\subsection{H1: The LLM Score Captures Real Economic Content}
\label{sec:results_h1}

The first hypothesis asks whether the frozen LLM score contains economically meaningful information about future firm outcomes. Using the primary model (GPT-4.1, June~2024 cutoff), we regress three post-cutoff fundamental and analyst variables on the sector-neutral LLM z-score~$\tilde{S}_{i,m}$, controlling for size, quality, value and momentum factor exposures. The sample comprises all firms with non-missing LLM scores and factor data ($N = 6{,}600$); analyst target-price coverage is lower, reducing the target-price sample to $N = 5{,}057$.

Table~\ref{tab:h1} reports the results. The LLM score is a strong predictor of realised twelve-month revenue growth ($\beta = 0.168$, $t = 10.89$), suggesting that the model's textual assessment anticipates the trajectory of top-line fundamentals. The score also predicts realised net-income growth ($\beta = 0.038$, $t = 2.18$) and analyst target-price revisions ($\beta = 0.347$, $t = 4.80$), both significant at conventional levels. All three coefficients carry the expected positive sign: firms that the LLM rates more favourably subsequently experience stronger fundamental performance and upward analyst revisions.

\begin{table}[htbp]
\centering
\small
\sbox0{%
\begin{tabular}{lccc}
\toprule
 & $\Delta\text{Revenue}_{12\text{m}}$
 & $\Delta\text{Net Income}_{12\text{m}}$
 & $\Delta\text{Target Price}_{12\text{m}}$ \\
\midrule
$\tilde{S}^{\text{LLM}}$ & $0.168^{***}$ & $0.038^{**}$ & $0.347^{***}$ \\
 & $(10.89)$ & $(2.18)$ & $(4.80)$ \\[6pt]
Controls & Yes & Yes & Yes \\
$N$ & 6{,}600 & 6{,}600 & 5{,}057 \\
\bottomrule
\end{tabular}}%
\begin{minipage}{\wd0}
\caption{H1 validation: LLM score predicts post-cutoff fundamentals and
analyst revisions (GPT-4.1).}
\label{tab:h1}
\usebox0
\par\smallskip
\footnotesize
$t$-statistics in parentheses. Controls include $z$-scored log market capitalisation, quality, value and momentum factor composites (four controls only; the eight-control H2 specification additionally requires cheapness metrics, yielding a smaller sample). The H1 sample ($N = 6{,}600$) is therefore larger than the H2 regression-ready sample in Table~\ref{tab:sample_construction}. Target-price coverage is lower, reducing the third column to $N = 5{,}057$. All variables are sector-neutral z-scores. $^{***}$, $^{**}$, $^{*}$ denote significance at the 1\%, 5\% and 10\% levels, respectively.
\end{minipage}
\end{table}

These results suggest that the LLM outlook score is not noise: it encodes information about the subsequent trajectory of firm fundamentals that analysts will later incorporate into their forecasts. This validation is a necessary precondition for interpreting the outlook score's return predictability as economically meaningful. If the LLM score failed to predict real outcomes, any return predictability would be uninformative and likely reflect data-mining rather than genuine information content.

\subsection{H2: The LLM Outlook Score Predicts Cross-Sectional Returns}
\label{sec:results_h2}

The core test follows the \cite{fama1973risk} cross-sectional regression approach, regressing post-cutoff cumulative excess returns on the sector-neutral LLM outlook score~$\tilde{S}_{i,t}^{\text{LLM}}$, controlling for size, quality, value, momentum and four market-implied cheapness metrics (inverse P/E, ICE-GLS, ICE-PEG and inverse EV/EBITDA).

\subsubsection{Single-Model Results}

Table~\ref{tab:h2_single} reports the cross-sectional regression results for the primary workhorse model (GPT-4.1, knowledge cutoff June~2024). At the one-month horizon, the outlook-score coefficient is $\gamma = 0.0122$ with $t = 4.25$, significant at the 1\% level. A one-standard-deviation increase in the outlook score---meaning the LLM views the firm one standard deviation more favourably---predicts a 1.22 percentage-point higher return over the subsequent month, after controlling for what market prices already imply about relative valuation.

The coefficient remains economically and statistically significant at longer horizons: $t = 3.28$ at six months and $t = 4.15$ at twelve months. The intermediate horizons are weaker ($t = 0.63$ at three months, $t = 0.29$ at four months). This non-monotonicity likely reflects noise amplified by the single-cutoff design rather than a systematic pattern, as the signal re-emerges strongly at six and twelve months. The cumulative coefficient at twelve months ($\gamma = 0.068$) is over five times the one-month estimate, suggesting that the predictive content of the outlook score is persistent rather than transient.

\begin{table}[htbp]
\centering
\small
\sbox0{%
\begin{tabular}{lccccc}
\toprule
 & $\tau = 1$m & $\tau = 3$m & $\tau = 4$m
 & $\tau = 6$m & $\tau = 12$m \\
\midrule
$\gamma$ & $0.0122^{***}$ & $0.0038$ & $0.0020$
          & $0.0386^{***}$ & $0.0683^{***}$ \\
$t$-stat  & $4.25$ & $0.63$ & $0.29$ & $3.28$ & $4.15$ \\[4pt]
$R^2$     & $0.062$ & $0.017$ & $0.008$ & $0.026$ & $0.025$ \\
$N$       & 1{,}272 & 1{,}272 & 1{,}272 & 1{,}272 & 1{,}272 \\
\bottomrule
\end{tabular}}%
\begin{minipage}{\wd0}
\caption{H2 single-model test: LLM outlook score predicts
cross-sectional returns (GPT-4.1).}
\label{tab:h2_single}
\usebox0
\par\smallskip
\footnotesize
OLS cross-sectional regressions of cumulative excess returns on the sector-neutral LLM outlook score, with eight controls: size (log market capitalisation), quality, value and momentum factor composites, and four market-implied cheapness metrics (inverse P/E, ICE-GLS, ICE-PEG and inverse EV/EBITDA). All regressors are sector-neutral z-scores. $^{***}\,p<0.01$; $^{**}\,p<0.05$; $^{*}\,p<0.10$.
\end{minipage}
\end{table}

\subsubsection{Incremental Predictive Content: LLM Score versus Cheapness}

A key question is whether the outlook score's predictive power is genuinely textual or simply proxies for market-implied cheapness. We address this by estimating four nested specifications at the one-month horizon, all on a common sample with factor-only controls (excluding cheapness controls to avoid mechanical collinearity): (i)~returns on the composite cheapness z-score alone (average of the four market metrics); (ii)~returns on the LLM outlook z-score alone; (iii)~returns on their additive mismatch ($\tilde{S} + \tilde{C}$); and (iv)~an unrestricted specification with both entered separately. The results are striking. The LLM-only specification ($t = 4.96$, $R^2 = 0.044$) dominates the cheapness-only specification ($R^2 = 0.043$) and the additive mismatch ($R^2 = 0.042$). In the unrestricted model ($R^2 = 0.060$), the LLM score enters positively and significantly ($\gamma = 0.0133$, $t = 4.74$) while the cheapness composite enters with a \emph{negative} coefficient ($\delta = -0.251$, $t = -4.69$) conditional on the controls, reflecting the fact that the standard value control already absorbs the value premium and renders residual cheapness a contrarian signal. A Wald test decisively rejects the equal-weight restriction ($F = 24.43$, $p < 0.001$), indicating that the additive mismatch is a mis-specified combination of the two components.

These results motivate our decision to use the LLM outlook score as the primary predictor rather than the composite mismatch. The predictive content appears to reside in the qualitative textual signal---management tone, competitive positioning, strategic outlook---rather than in its interaction with accounting-based valuation. The cheapness metrics remain in the regression as controls, ensuring that the outlook-score coefficient is estimated conditional on what market prices already imply.

\subsubsection{Pooled Panel with Model Fixed Effects}

To exploit variation across all twelve model checkpoints simultaneously, we stack the individual cross-sections into a pooled panel ($N = 14{,}918$ firm--model observations) and include model fixed effects to absorb checkpoint-specific intercepts. Standard errors are clustered at the model level.

Table~\ref{tab:h2_pooled} reports the results. The pooled outlook-score coefficient is $\gamma = 0.0074$ ($t = 6.02$) at one month and $\gamma = 0.0113$ ($t = 1.07$) at twelve months, indicating that the single-model result is not an artefact of one particular checkpoint. The coefficient is significant at all horizons up to six months, with the strongest result at one month.

\begin{table}[htbp]
\centering
\small
\sbox0{%
\begin{tabular}{lccccc}
\toprule
 & $\tau = 1$m & $\tau = 3$m & $\tau = 4$m
 & $\tau = 6$m & $\tau = 12$m \\
\midrule
\multicolumn{6}{l}{\emph{Panel A: Model fixed effects}} \\[2pt]
$\gamma$ & $0.0074^{***}$ & $0.0082^{***}$ & $0.0074^{***}$
          & $0.0119^{**}$ & $0.0113$ \\
$t$-stat  & $6.02$ & $3.08$ & $3.28$ & $2.30$ & $1.07$ \\[2pt]
$R^2$     & $0.021$ & $0.010$ & $0.009$ & $0.013$ & $0.015$ \\
$N$       & 14{,}918 & 14{,}918 & 14{,}918 & 14{,}918 & 12{,}752 \\[6pt]
\multicolumn{6}{l}{\emph{Panel B: Model + industry fixed effects
($\tau = 1$m)}} \\[2pt]
$\gamma$ & \multicolumn{5}{c}{$0.0064^{***}\quad (t = 5.74)$} \\
Sector FE & \multicolumn{5}{c}{11 GICS sectors} \\
$R^2$     & \multicolumn{5}{c}{$0.045$} \\
$N$       & \multicolumn{5}{c}{14{,}918} \\
\bottomrule
\end{tabular}}%
\begin{minipage}{\wd0}
\caption{H2 pooled panel: LLM outlook score predicts returns across model checkpoints.}
\label{tab:h2_pooled}
\usebox0
\par\smallskip
\footnotesize
Panel~A reports pooled OLS regressions with model fixed effects and standard errors clustered at the model level. Panel~B adds GICS sectors fixed effects to the one-month specification. The outlook-score coefficient is virtually unchanged, indicating that the effect is not driven by sector-level omitted variables. $^{***}\,p<0.01$; $^{**}\,p<0.05$; $^{*}\,p<0.10$.
\end{minipage}
\end{table}

Panel~B of Table~\ref{tab:h2_pooled} adds GICS sector fixed effects to the one-month specification. Despite the sector-neutral construction of the outlook score, one might worry that residual industry effects drive the result. The coefficient is virtually unchanged at $\gamma = 0.0064$ ($t = 5.74$, $R^2 = 0.045$), suggesting that the outlook score captures firm-specific information content beyond any industry tilt.

\subsubsection{Fama--MacBeth Style Cross-Model Inference}

As an alternative to the pooled panel, we treat each model checkpoint as an independent cross-section and compute the average outlook-score coefficient across the twelve models. The mean $\gamma = 0.0078$ with an i.i.d.\ $t$-statistic of~$5.09$ (12~cross sections); adjusting for possible autocorrelation with a Newey--West kernel (one lag) yields $t = 4.71$ (Panel~C of Table~\ref{tab:overlap_robust}). A non-parametric bootstrap (10{,}000~resamples) produces a 95\% confidence interval of $[0.0047,\; 0.0103]$ that excludes zero, with a bootstrap $p$-value of~$0.0002$.

The twelve models map to only six unique analysis dates, with six models sharing the May/June~2024 cluster.\footnote{Four models (GPT-4.1, GPT-4.1-mini, GPT-4.1-nano, o3) have an official cutoff of June~2024, while two (GPT-5-mini, GPT-5-nano) have a cutoff of May~2024. Because returns are measured from the last trading day on or before the cutoff date, both groups map to the same analysis date (last trading day of May~2024). The six unique analysis dates are therefore September~2021, October~2023, December~2023, May~2024, September~2024 and August~2025.} For any given firm, the post-cutoff return is identical across all models within a cutoff cluster, creating overlapping observations. This raises a legitimate concern about the independence assumptions underlying the pooled-panel inference. We address this concern through three complementary approaches.

Table~\ref{tab:overlap_robust} summarises the three approaches.

\paragraph{Driscoll--Kraay standard errors.}
The most direct correction retains all twelve models but computes standard errors that are robust to arbitrary cross-sectional dependence, temporal dependence and heteroskedasticity \citep{driscoll1998consistent}. Grouping moment conditions by the six unique cutoff dates and applying a Newey--West kernel (one lag), the outlook-score coefficient remains significant at all horizons (Panel~A). The information-environment interactions (Section~\ref{sec:results_h3}) are equally robust: the analyst-coverage interaction $t$-statistic rises to~$3.37$ (from~$2.44$) and price-target coverage to~$3.19$ (from~$3.02$).

\paragraph{One-model-per-cutoff Fama--MacBeth.}
As a stricter test, we eliminate overlapping observations entirely by selecting only the largest model at each cutoff date, yielding six independent cross-sections. We average the resulting $\gamma$ coefficients following the Fama--MacBeth procedure, with Newey--West adjustment (one lag). The one-month result is marginally significant ($p = 0.074$), reflecting the low power of six time-series observations, but the bootstrap $p$-value is~$0.019$ and the effect strengthens at longer horizons (Panel~B).

\paragraph{Twelve-model Fama--MacBeth.}
Treating each of the twelve checkpoints as a cross-section overstates significance because models within a cutoff cluster share the same return realisation. Nevertheless, the Newey--West $t$-statistics remain significant through six months and the bootstrap $p$-values are below~$0.05$ at all horizons except twelve months (Panel~C), consistent with the more conservative tests.

\begin{table}[htbp]
\centering
\small
\sbox0{%
\begin{tabular}{llccccc}
\toprule
& & \multicolumn{5}{c}{Return horizon} \\
\cmidrule(lr){3-7}
& & 1\,m & 3\,m & 4\,m & 6\,m & 12\,m \\
\midrule
\multicolumn{7}{l}{\textit{Panel~A: Driscoll--Kraay SEs (12 models, $T_{\text{eff}} = 6$ cutoffs)}} \\[3pt]
& $\hat\gamma$    & $0.0074$ & $0.0082$ & $0.0074$ & $0.0119$ & $0.0113$ \\
& $t_{\text{DK}}$ & $5.43^{***}$ & $4.60^{***}$ & $4.12^{***}$ & $3.84^{**}$ & $2.19^{*}$ \\
& $t_{\text{model}}$ & $6.02$ & $3.08$ & $3.28$ & $2.30$ & $1.07$ \\[6pt]
\multicolumn{7}{l}{\textit{Panel~B: One-model-per-cutoff FM + Newey--West ($T = 6$)}} \\[3pt]
& $\bar\gamma$    & $0.0064$ & $0.0142$ & $0.0138$ & $0.0159$ & $0.0424$ \\
& $t_{\text{NW}}$ & $2.25^{*}$ & $2.81^{**}$ & $3.20^{**}$ & $2.45^{*}$ & $2.45^{*}$ \\
& $p_{\text{boot}}$ & $0.019$ & $0.015$ & $0.000$ & $0.036$ & $0.000$ \\[6pt]
\multicolumn{7}{l}{\textit{Panel~C: Twelve-model FM + Newey--West ($T = 12$)}} \\[3pt]
& $\bar\gamma$       & $0.0078$ & $0.0100$ & $0.0084$ & $0.0123$ & $0.0158$ \\
& $t_{\text{NW}}$    & $4.71^{***}$ & $3.35^{***}$ & $3.99^{***}$ & $2.26^{**}$ & $1.46$ \\
& $p_{\text{boot}}$  & $0.000$ & $0.002$ & $0.003$ & $0.021$ & $0.204$ \\
\bottomrule
\end{tabular}}%
\begin{minipage}{\wd0}
\caption{Robustness to overlapping observations.}
\label{tab:overlap_robust}
\usebox0
\par\smallskip
\footnotesize
Panel~A applies Driscoll--Kraay standard errors to the full twelve-model pooled panel, grouping moment conditions by cutoff date (Newey--West kernel, one lag). Panel~B selects the largest model at each of six cutoff dates, runs cross-sectional regressions and averages coefficients \`a la Fama--MacBeth, with Newey--West adjustment for autocorrelation. Panel~C treats all twelve checkpoints as cross-sections with Newey--West adjustment; significance may be overstated due to within-cutoff overlap. $t_{\text{model}}$ denotes model-clustered $t$-statistics from the pooled panel for comparison. $^{***}\,p<0.01$; $^{**}\,p<0.05$; $^{*}\,p<0.10$.
\end{minipage}
\end{table}

We note two further econometric caveats. First, the ``Fama--MacBeth style'' label is used loosely: the cross-sections are defined by model checkpoints (or cutoff dates) rather than calendar periods, so the averaging does not address the same time-series dependence that the classical procedure targets. Second, the pooled-panel standard errors clustered at the model level rest on only twelve clusters, below the threshold (typically 30--50) at which cluster-robust inference is reliable \citep{cameron2008bootstrap}. The Driscoll--Kraay and one-model-per-cutoff Fama--MacBeth procedures both sidestep this limitation by treating the six cutoff dates as the effective time dimension.

Taken together, the single-model results ($t = 4.25$), the Driscoll--Kraay pooled panel ($t = 5.43$), the clean Fama--MacBeth procedure (bootstrap $p = 0.019$) and the horizon structure (Newey--West $t > 2.4$ at three, four and twelve months) consistently show that the LLM outlook score predicts cross-sectional returns after controlling for market-implied valuations. The result is not an artefact of overlapping observations.

\subsection{H3: Cross-Sectional Heterogeneity}
\label{sec:results_h3}

Given that the outlook score predicts returns on average, we ask whether the effect varies systematically with firm characteristics. We examine two dimensions: firm size and the information environment.

\subsubsection{Size-Based Subsample Analysis}

We sort firms into terciles by market capitalisation and estimate the outlook-score regression separately within each group. Table~\ref{tab:h3_size} reports the results. The outlook-score effect is statistically significant in the small and large terciles: small firms ($\gamma = 0.0149$, $t = 2.18$), mid-cap firms ($\gamma = 0.0077$, $t = 1.38$) and large firms ($\gamma = 0.0087$, $t = 2.40$). The formal interaction between the outlook score and log market capitalisation is marginally significant ($t = 1.88$, $p = 0.06$), suggesting a positive size interaction though the pattern across terciles is non-monotonic.

\begin{table}[htbp]
\centering
\small
\sbox0{%
\begin{tabular}{lccc}
\toprule
 & Small & Mid & Large \\
\midrule
$\gamma$  & $0.0149^{**}$ & $0.0077$ & $0.0087^{**}$ \\
$t$-stat   & $2.18$ & $1.38$ & $2.40$ \\[4pt]
$N$        & 424 & 424 & 424 \\
\bottomrule
\end{tabular}}%
\begin{minipage}{\wd0}
\caption{H3 size terciles: Outlook-score--return relation by firm size
(GPT-4.1, $\tau = 1$m).}
\label{tab:h3_size}
\usebox0
\par\smallskip
\footnotesize
Firms sorted into terciles by log market capitalisation within the regression sample. Controls include market-implied cheapness metrics and factor exposures as in Table~\ref{tab:h2_single}. $^{***}\,p<0.01$; $^{**}\,p<0.05$; $^{*}\,p<0.10$.
\end{minipage}
\end{table}

\subsubsection{Information Environment Interactions}

We interact the outlook score with five proxies for the firm's information environment: analyst coverage (log number of analysts), price-target coverage, analyst price-target dispersion (a standard disagreement proxy), financial leverage (debt-to-assets) and average daily trading volume. Table~\ref{tab:h3_interactions} reports the interaction coefficients from both the single-model specification and the pooled panel with model fixed effects.

\begin{table}[htbp]
\centering
\small
\sbox0{%
\begin{tabular}{lcccc}
\toprule
& \multicolumn{2}{c}{Single-model (GPT-4.1)}
& \multicolumn{2}{c}{Pooled panel} \\
\cmidrule(lr){2-3} \cmidrule(lr){4-5}
Interaction variable & $\beta_{\text{int}}$ & $t$-stat
                     & $\beta_{\text{int}}$ & $t$-stat \\
\midrule
Analyst coverage     & $0.0040$ & $1.58$ & $0.0026^{**}$ & $2.44$ \\
PT coverage          & $0.0028$ & $1.19$ & $0.0030^{***}$ & $3.02$ \\
PT dispersion        & $-0.0029$ & $-0.68$ & $0.0026$ & $1.52$ \\
Debt/Assets          & $-0.0116$ & $-1.59$ & $-0.0015$ & $-0.34$ \\
Trading volume       & $0.0000$ & $1.18$ & $0.0000$ & $0.53$ \\
\bottomrule
\end{tabular}}%
\begin{minipage}{\wd0}
\caption{H3 information-environment interactions ($\tau = 1$m).}
\label{tab:h3_interactions}
\usebox0
\par\smallskip
\footnotesize
Each row reports the coefficient on $\tilde{S}_{i,t}^{\text{LLM}} \times z_{i,t}^{\text{proxy}}$ from a regression that also includes the main outlook-score effect and controls (including market-implied cheapness metrics). The pooled panel includes model fixed effects with standard errors clustered at the model level. $^{***}\,p<0.01$; $^{**}\,p<0.05$; $^{*}\,p<0.10$.
\end{minipage}
\end{table}

In the single-model specification, analyst coverage is marginally significant ($t = 1.58$) and leverage has a negative sign ($t = -1.59$), but most interactions lack power with approximately 1,200 observations. The pooled panel reveals a clearer pattern: the outlook-score effect is significantly stronger for firms with higher analyst coverage ($t = 2.44$) and price-target coverage ($t = 3.02$). The positive sign on the coverage interactions may appear surprising, but is consistent with the interpretation that analyst-covered firms generate richer textual records---more reports, filings and commentary---from which the LLM can extract nuanced information. Leverage and trading volume interactions are insignificant in the pooled specification.

The baseline outlook-score coefficient remains strongly significant even after accounting for variation in the information environment, suggesting a general informational friction amplified modestly when the public record is richer.

\subsection{H4: Price Convergence and Model Sophistication}
\label{sec:results_h4}

\subsubsection{Cumulative Price Convergence}

If the outlook score captures temporarily unpriced information, the predictive coefficient should grow with the return horizon as prices gradually converge toward the LLM-implied valuation. We test this by computing the Spearman rank correlation between the return horizon $\tau \in \{1, 3, 4, 6, 12\}$ months and the corresponding outlook-score coefficient~$\gamma_{\tau}$.

For the primary model (GPT-4.1), the rank correlation is $r = 0.910$ ($p = 0.032$); in the pooled panel, $r = 0.770$ ($p = 0.128$). The positive and significant single-model correlation is consistent with prices moving \emph{toward} the LLM's assessment over time: the outlook score becomes a stronger predictor at longer horizons, consistent with gradual information incorporation and limits to arbitrage \citep{hong1999unified,shleifer1997limits}. The pooled-panel correlation is positive but not significant at conventional levels, reflecting the noisier horizon structure when averaging across heterogeneous checkpoints.

\begin{figure}[htbp]
\centering
\includegraphics[width=\textwidth]{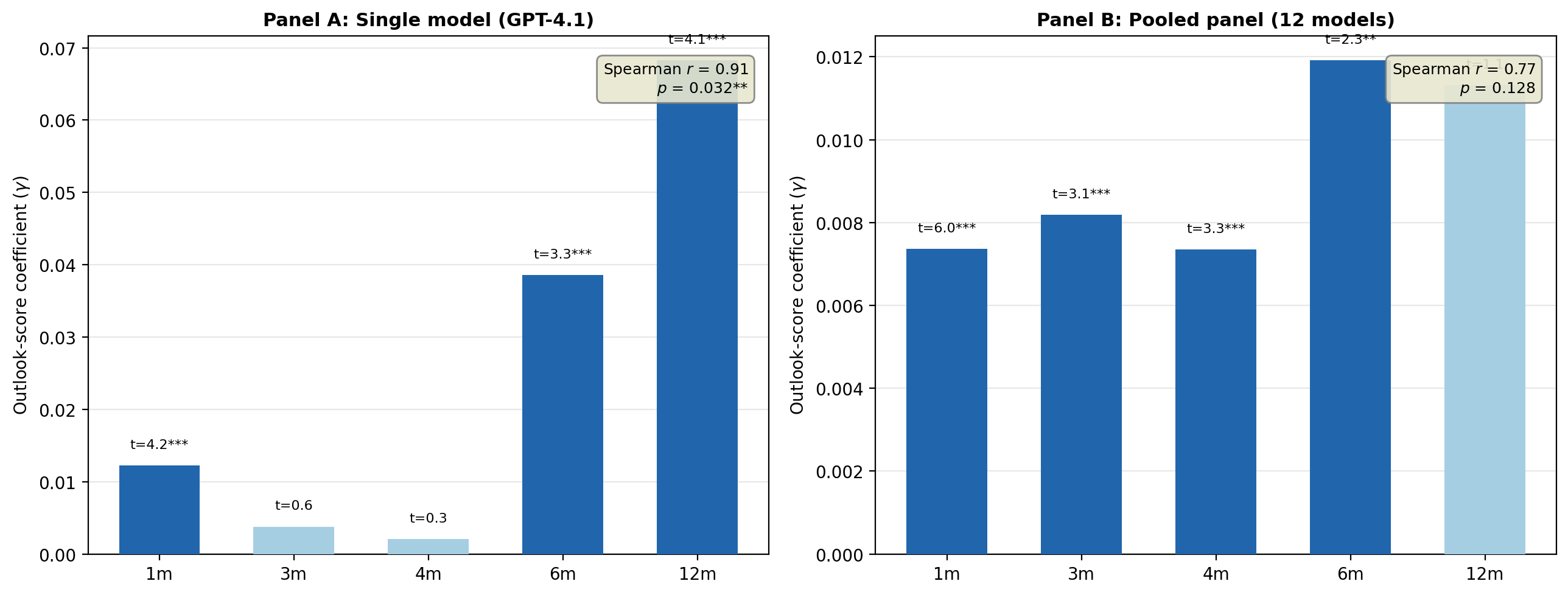}
\caption{Outlook-score coefficient $\gamma$ across return horizons. Panel~A shows the single-model specification (GPT-4.1); Panel~B shows the pooled panel with model fixed effects. Dark bars indicate significance at the 5\% level. Despite an intermediate-horizon dip in Panel~A, the overall trend is positive and significant (Spearman rank correlations reported in the insets).}
\label{fig:convergence}
\end{figure}

This growth pattern contrasts with a transient-sentiment interpretation, under which the coefficient would peak at short horizons and then decay. The pattern instead is consistent with slow fundamental convergence. Figure~\ref{fig:convergence} visualises the horizon structure for both specifications.

\subsubsection{Cross-Model Comparison}

Table~\ref{tab:h4_models} reports the outlook-score coefficient at $\tau = 1$ month for each of the twelve model checkpoints. The estimates range from $-0.005$ (GPT-4o, cutoff October~2023) to $0.015$ (GPT-5.4), with eleven of twelve models producing positive point estimates and seven reaching statistical significance at the 5\% level. The strongest results are concentrated in the May/June~2024 cutoff cluster, where three models exceed $t > 3$ (GPT-4.1, GPT-4.1-mini, GPT-4.1-nano) and a fourth approaches it (o3, $t = 2.96$).

\begin{table}[htbp]
\centering
\small
\sbox0{%
\begin{tabular}{llrrc}
\toprule
Model & Cutoff & $\gamma$ & $t$-stat & $N$ \\
\midrule
GPT-3.5-Turbo & Sep 2021  & $0.0017$ & $0.63$ & 1{,}448 \\
GPT-4o        & Oct 2023  & $-0.0047$ & $-0.95$ & 1{,}059 \\
GPT-4-Turbo   & Dec 2023  & $0.0085^{*}$ & $1.69$ & 1{,}208 \\
\addlinespace
GPT-5-mini    & May 2024  & $0.0092^{***}$ & $2.59$ & 1{,}267 \\
GPT-5-nano    & May 2024  & $0.0059^{**}$ & $2.09$ & 1{,}271 \\
GPT-4.1       & Jun 2024  & $0.0122^{***}$ & $4.25$ & 1{,}272 \\
GPT-4.1-mini  & Jun 2024  & $0.0108^{***}$ & $4.01$ & 1{,}272 \\
GPT-4.1-nano  & Jun 2024  & $0.0065^{***}$ & $3.11$ & 1{,}272 \\
o3            & Jun 2024  & $0.0113^{***}$ & $2.96$ & 1{,}272 \\
\addlinespace
GPT-5         & Sep 2024  & $0.0059$ & $1.22$ & 1{,}411 \\
GPT-5.2       & Aug 2025  & $0.0109^{*}$ & $1.69$ & 1{,}083 \\
GPT-5.4       & Aug 2025  & $0.0149^{**}$ & $2.01$ & 1{,}083 \\
\bottomrule
\end{tabular}}%
\begin{minipage}{\wd0}
\caption{Cross-model outlook-score coefficients ($\tau = 1$m).}
\label{tab:h4_models}
\usebox0
\par\smallskip
\footnotesize
Each row reports the OLS coefficient on the sector-neutral LLM outlook score from a separate cross-sectional regression of one-month excess returns, with the same control set as Table~\ref{tab:h2_single}. Models ordered by knowledge cutoff date.
\end{minipage}
\end{table}

\begin{figure}[htbp]
\centering
\includegraphics[width=\textwidth]{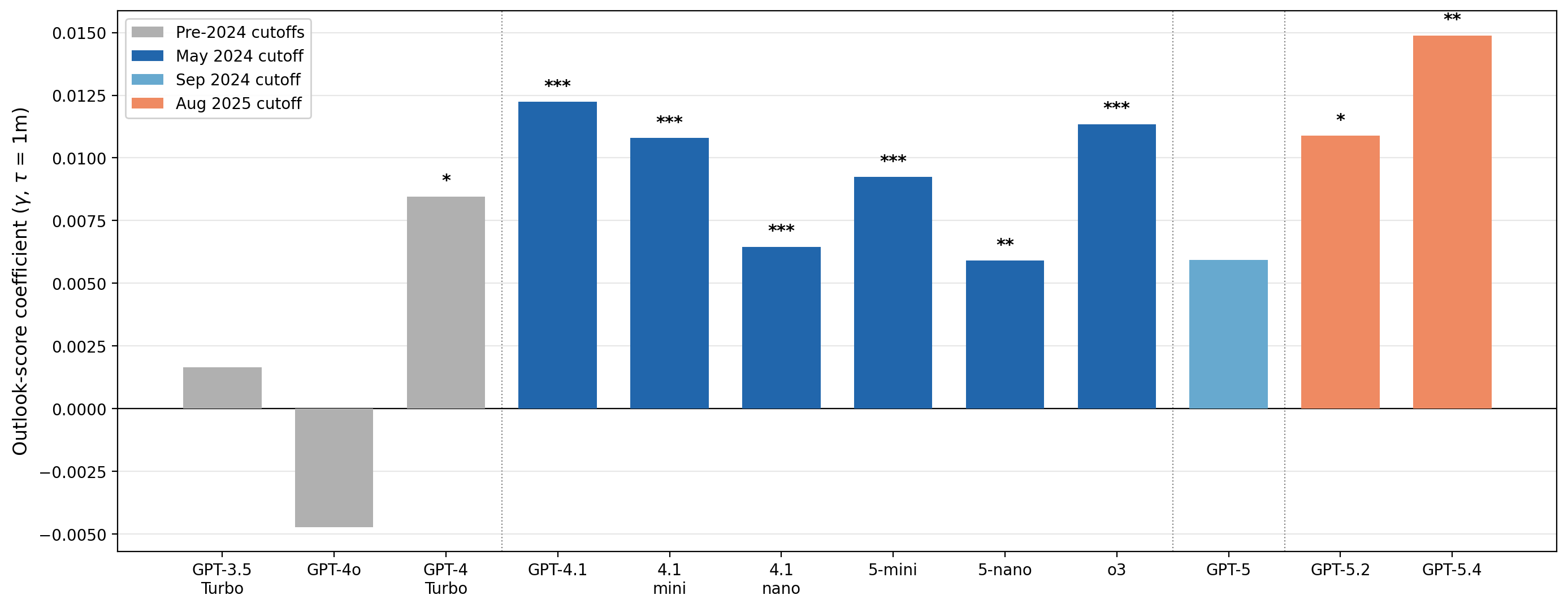}
\caption{Outlook-score coefficient $\gamma$ at $\tau = 1$ month for each of the twelve model checkpoints. Bars are colour-coded by knowledge cutoff date. Black outlines indicate significance at the 5\% level. Dashed vertical lines separate cutoff clusters.}
\label{fig:cross_model}
\end{figure}

Figure~\ref{fig:cross_model} plots the cross-model results, colour-coded by knowledge cutoff date. The May/June~2024 cluster dominates the significant results and all but one model produce positive point estimates.

\subsubsection{Model Sophistication}

The Spearman correlation between cutoff recency and $\gamma$ is $0.617$ ($p = 0.033$), indicating a statistically significant positive relationship: newer models tend to produce larger coefficients. The correlation with model generation is $0.489$ ($p = 0.106$), positive but not significant. Notably, within the May/June~2024 cutoff cluster, larger models produce stronger coefficients (GPT-4.1: $0.0122$ vs.\ GPT-4.1-nano: $0.0065$; o3: $0.0113$ vs.\ GPT-5-nano: $0.0059$), suggesting that model capability amplifies the signal even when the information set is held constant.

\subsection{Portfolio-Based Evidence}
\label{sec:results_portfolio}

To complement the regression evidence, we sort firms into quintiles by the LLM outlook score and examine mean one-month returns. We report two sets of portfolio sorts: an unconditional sort on the raw outlook score across the full sample, and a control-adjusted sort that first orthogonalises the outlook score with respect to the regression controls.

Table~\ref{tab:portfolio} reports the unconditional results for the primary model (GPT-4.1), with returns winsorised at the 1st/99th percentiles. The long-short portfolio (Q5~$-$~Q1) earns a mean return of $5.95\%$ per month ($t = 7.53$). The return pattern is monotonically increasing across quintiles, from $-6.56\%$ in Q1 to $-0.61\%$ in Q5. The unconditional sort does not control for factors or cheapness, so these raw spreads partly reflect factor exposures.

More informative is the control-adjusted sort, which residualises the outlook score on the same controls used in the cross-sectional regression. The control-adjusted long-short spread is $2.55\%$ per month ($t = 3.71$), corresponding to an annualised return of approximately $30.6\%$. This suggests that the outlook score's economic content survives the removal of factor and cheapness exposures.

\begin{table}[htbp]
\centering
\small
\sbox0{%
\begin{tabular}{lrrr}
\toprule
Quintile & Mean return (\%) & Std.\ dev.\ & $N$ \\
\midrule
Q1 (most negative $\tilde{S}$) & $-6.56$ & $0.269$ & 1{,}330 \\
Q2 & $-2.91$ & $0.154$ & 1{,}336 \\
Q3 & $-2.51$ & $0.130$ & 1{,}322 \\
Q4 & $-2.20$ & $0.105$ & 1{,}334 \\
Q5 (most positive $\tilde{S}$) & $-0.61$ & $0.102$ & 1{,}325 \\
\midrule
L/S (Q5 $-$ Q1) & $+5.95$ & & \\
$t$-stat & $7.53$ & & \\
\bottomrule
\end{tabular}}%
\begin{minipage}{\wd0}
\caption{Unconditional portfolio sorts by LLM outlook score (GPT-4.1, $\tau = 1$m).}
\label{tab:portfolio}
\usebox0
\par\smallskip
\footnotesize
Firms sorted into quintiles by the sector-neutral LLM outlook score $\tilde{S}_{i,t}^{\text{LLM}}$ across all firms with non-missing scores and returns ($N = 6{,}647$; this exceeds the regression-ready sample in Table~\ref{tab:sample_construction} because the sort does not require non-missing cheapness controls). Returns winsorised at the 1st/99th percentiles. Mean one-month cumulative excess returns reported in percentage points. All quintile means are negative because the GPT-4.1 cutoff (June~2024) coincides with a broad market decline in the subsequent month; the long-short spread captures the relative difference across quintiles and is unaffected by the market level. The control-adjusted long-short spread (residualising the outlook score on regression controls) is $2.55\%$ ($t = 3.71$), annualised $30.6\%$.
\end{minipage}
\end{table}

\subsubsection{Factor Exposures}

The twelve model checkpoints map to only six unique cutoff dates (calendar months), since six models share the May/June~2024 cutoff cluster. Averaging across models within each analysis month yields six calendar-month long-short returns. With six observations and seven parameters (intercept plus FF5+Momentum), a formal time-series alpha regression is not feasible (negative degrees of freedom). We therefore report factor exposures descriptively rather than estimating alpha. The cross-sectional regressions in Section~\ref{sec:results_h2}---which control directly for size, value, quality and momentum with $N > 1{,}000$---remain the appropriate test of risk-adjusted predictability.

Despite this limitation, the factor structure of the long-short portfolio is informative about what the LLM outlook score captures. Correlating the six calendar-month long-short returns with contemporaneous factor returns, the strongest loading is on profitability (RMW, $\rho = +0.63$), followed by investment (CMA, $\rho = -0.52$) and size (SMB, $\rho = -0.57$). The market correlation is also negative ($\rho = -0.50$). High-outlook-score firms---those the LLM views most favourably---thus tend to be profitable and capital-light, consistent with the LLM recognising operational quality not yet fully reflected in prices relative to our controls. The portfolio tilts toward larger firms and away from the market factor, a profile broadly consistent with quality investing. Momentum exposure is near zero ($\rho = +0.16$), suggesting the signal is distinct from trend-following strategies. A simple CAPM regression---the only factor model feasible with four residual degrees of freedom---yields an insignificant intercept of $1.33\%$ per month ($t = 0.78$), underscoring that six calendar months provide insufficient power for time-series inference. These exposures are consistent with the cross-sectional regression controls: the outlook-score coefficient survives direct controls for quality, value and momentum factors with $N > 1{,}000$, indicating that the textual signal contains information beyond these standard dimensions.

\subsection{Robustness}
\label{sec:results_robustness}

We subject the core H2 result to a series of robustness checks.

\subsubsection{Micro-Cap Exclusion}

A common concern in cross-sectional return prediction is that results are driven by the smallest, least liquid firms. We re-estimate the baseline regression after excluding the bottom 20\% of firms by market capitalisation within the sample. The outlook-score coefficient \emph{increases} to $\gamma = 0.0133$ ($t = 4.42$, $N = 1{,}017$). Restricting further to the largest 50\% of firms (large-cap only), the coefficient remains significant at $\gamma = 0.0110$ ($t = 3.26$, $N = 636$). The outlook-score effect is therefore not a micro-cap phenomenon.

\subsubsection{Return Winsorisation}

Winsorising both returns and the outlook score at the 1st/99th percentiles yields $\gamma = 0.0125$ ($t = 4.51$); at the more aggressive 5th/95th percentiles, the coefficient is $\gamma = 0.0118$ ($t = 4.53$). The stability of the estimate under increasingly aggressive tail trimming indicates that the result is not driven by outliers.

\subsubsection{Alternative Mismatch Specifications (Robustness)}

As a robustness exercise, we also construct additive mismatch composites $M_{i,t}^{k} = \tilde{S}_{i,t}^{\text{LLM}} + \tilde{C}_{i,t}^{k}$ for each cheapness dimension~$k$. Table~\ref{tab:robustness_legs} reports these results. All four individual mismatch legs produce positive coefficients and three reach statistical significance: inverse P/E ($\beta = 0.0210$, $t = 5.72$), ICE-GLS ($\beta = 0.0094$, $t = 4.80$) and inverse EV/EBITDA ($\beta = 0.0097$, $t = 2.39$). The ICE-PEG leg is positive but not significant ($\beta = 0.0038$, $t = 1.07$). The equal-weighted composite ($\beta = 0.0133$, $t = 4.71$) lies between the individual legs, as expected.

\begin{table}[htbp]
\centering
\small
\sbox0{%
\begin{tabular}{lrrl}
\toprule
Mismatch leg & $\beta$ & $t$-stat & \\
\midrule
$M^{\text{ICE-GLS}}$     & $0.0094$ & $4.80$ & $^{***}$ \\
$M^{\text{InvPE}}$       & $0.0210$ & $5.72$ & $^{***}$ \\
$M^{\text{ICE-PEG}}$     & $0.0038$ & $1.07$ & \\
$M^{\text{InvEVEB}}$     & $0.0097$ & $2.39$ & $^{**}$ \\[4pt]
$M^{\text{Composite}}$   & $0.0133$ & $4.71$ & $^{***}$ \\
\bottomrule
\end{tabular}}%
\begin{minipage}{\wd0}
\caption{Robustness: Individual and composite mismatch specifications ($\tau = 1$m).}
\label{tab:robustness_legs}
\usebox0
\par\smallskip
\footnotesize
Each row reports a separate regression of one-month excess returns on the indicated mismatch measure ($M^k = \tilde{S}^{\text{LLM}} + \tilde{C}^k$), with factor-only controls (size, quality, value, momentum---excluding the cheapness controls already embedded in the mismatch). Sample sizes vary by cheapness measure availability: $N = 1{,}480$ (ICE-GLS), $N = 5{,}280$ (InvPE), $N = 4{,}404$ (ICE-PEG), $N = 4{,}568$ (InvEVEB), $N = 1{,}272$ (Composite).
\end{minipage}
\end{table}

We verify sign-convention consistency by examining pairwise correlations among the four mismatch legs. The average correlation between $M^{\text{InvPE}}$ and the other three legs is~$0.77$ and all six pairwise correlations are positive (range: $0.51$--$0.82$), consistent with all legs sharing a common factor driven by the LLM component. The inverse P/E leg is in fact the strongest individual mismatch predictor, but as shown in the incremental decomposition, this predictive power is attributable to the LLM score rather than to the cheapness component.

\subsubsection{Summary of Robustness Checks}

Table~\ref{tab:robustness_summary} summarises the full battery of robustness tests.

\begin{table}[htbp]
\centering
\small
\sbox0{%
\begin{tabular}{lrrr}
\toprule
Specification & $\gamma$ & $t$-stat & $N$ \\
\midrule
Baseline                              & $0.0122$ & $4.25$ & 1{,}272 \\
Pooled panel (model FE)               & $0.0074$ & $6.02$ & 14{,}918 \\
Pooled panel (model + industry FE)    & $0.0064$ & $5.74$ & 14{,}918 \\
FM 12-model (bootstrap $p = 0.0002$)$^{\dagger}$ & $0.0078$ & $5.09$ & 12 \\
FM 6-cutoff + NW (bootstrap $p = 0.019$) & $0.0064$ & $2.25$ & 6 \\
Driscoll--Kraay SEs                   & $0.0074$ & $5.43$ & 14{,}918 \\
Excl.\ bottom 20\% by size           & $0.0133$ & $4.42$ & 1{,}017 \\
Large-cap only (top 50\%)            & $0.0110$ & $3.26$ & 636 \\
Winsorised 1/99                       & $0.0125$ & $4.51$ & 1{,}272 \\
Winsorised 5/95                       & $0.0118$ & $4.53$ & 1{,}272 \\
\bottomrule
\end{tabular}}%
\begin{minipage}{\wd0}
\caption{Summary of robustness checks ($\tau = 1$m, GPT-4.1).}
\label{tab:robustness_summary}
\usebox0
\par\smallskip
\footnotesize
All specifications use the LLM outlook score as the primary regressor and the same control set (including market-implied cheapness metrics) unless otherwise noted. Winsorisation caps extreme values at the indicated percentile rather than dropping observations, so $N$ is unchanged. The FM rows report the mean coefficient across model cross-sections; $t$-statistics are i.i.d.\ for the 12-model row and Newey--West adjusted for the 6-cutoff row. The Driscoll--Kraay row reports the pooled-panel coefficient with standard errors robust to cross-sectional and temporal dependence. $^{\dagger}$Less conservative due to within-cutoff overlap (see Table~\ref{tab:overlap_robust}).
\end{minipage}
\end{table}

The outlook-score--return relation is robust to the inclusion of industry fixed effects, the exclusion of small-cap firms, aggressive winsorisation, pooling across checkpoints and alternative inference procedures. The outlook-score $t$-statistic exceeds~$3.2$ in every within-sample specification; even the most conservative cross-model test (six independent cutoff dates, Newey--West adjustment) yields $t = 2.25$ with a bootstrap $p$-value of~$0.019$, and the Driscoll--Kraay pooled-panel $t$-statistic is~$5.43$.

\subsection{Portfolio Application}
\label{sec:portfolio}

As a practical application, we examine whether the outlook-score signal can be translated into a portfolio signal using regularised mean-variance optimisation. At each cutoff date, we map the sector-neutral outlook z-score~$\tilde{S}_{i,m}$ into an expected excess return via $\mu_i = \hat{\gamma}\,\tilde{S}_{i,m}$, where $\hat{\gamma} = 0.0074$ is the Driscoll--Kraay pooled estimate from the one-month horizon (see Table~\ref{tab:h2_pooled}). We use the one-month coefficient for all signal lives because the portfolio is rebalanced monthly: each month's optimisation forms a one-period expected-return input, regardless of how many months the signal remains active. The horizon structure documented in Section~\ref{sec:results_h4} implies that using a longer-horizon $\hat{\gamma}$ would scale up the expected return but also introduce compounding assumptions; the conservative one-month estimate avoids this.

The score entering the optimiser is the raw sector-neutral z-score~$\tilde{S}_{i,m}$, not the residual from the full regression that includes cheapness controls. This is deliberate: the regression decomposes predictability into a cheapness-correlated component and a residual textual component, but the portfolio is designed to capture the \emph{total} signal content of the outlook score as a standalone alpha input. Orthogonalising the score to cheapness before feeding it to the optimiser would discard the cheapness-correlated component and understate the score's practical usefulness. The trade-off is that the portfolio inherits any cheapness-factor tilt embedded in the raw score, so the reported returns overstate the signal's independent contribution relative to a strategy that already trades on valuation metrics. This caveat should be borne in mind when interpreting the results.

The covariance matrix is provided by an Axioma risk model, which decomposes stock-level risk into approximately 80~style and industry factors plus idiosyncratic variance. We apply light diagonal shrinkage to the factor covariance matrix, $\tilde{F} = (1-\alpha)\,F + \alpha\,\mathrm{diag}(F)$ with $\alpha = 0.10$; a well-estimated commercial factor model requires only modest regularisation. Outlook z-scores are winsorised at $\pm 2.5$ standard deviations before entering the optimiser to limit the influence of extreme assessments. The risk-aversion parameter is set to $\lambda = 1$, and we impose long-only constraints with a 1.5\% position cap (a conservative diversification norm that results in approximately 65--90 active positions per rebalance month). The portfolio is rebalanced monthly, the risk model is also updated each month while the alpha signal remains fixed from the cutoff date. All reported returns are gross of transaction costs. Because the alpha signal is fixed and only the risk model is updated, rebalancing turnover is moderate: the aggregate 12-month portfolio averages approximately 15\% one-way monthly turnover (excluding the initial formation month), reflecting the tighter position cap. At an assumed round-trip cost of 20~basis points, the aggregate 12-month portfolio's annualised return would decline by roughly 35~basis points per year (SR impact $\approx 0.05$); per-model portfolios would bear a cost of 50--80~basis points per year.

We consider three signal-life horizons (3, 6 and 12~months) and benchmark the mean-variance (MV) portfolio against both an equal-weight (EW) portfolio of the same universe and the S\&P~500 Index. GPT-3.5-Turbo is excluded from the portfolio sample: its September~2021 training cutoff precedes the next model (GPT-4o, October~2023) by over two years, creating a large temporal gap with no overlapping signal coverage. Including it conflates the 2022 bear-market period---where no other model provides a signal---with the denser post-2023 sample, introducing a composition effect that obscures the performance of the main model panel.

Table~\ref{tab:portfolio_summary} reports the results in two panels. Panel~A compares the aggregate portfolio across three signal-life horizons (3, 6 and 12~months), where the signal from the most capable model at each cutoff date is held for the indicated number of months before expiring. MV Sharpe ratios are 3.53 at 3~months, 2.33 at 6~months and 2.31 at 12~months, the non-monotonicity at the shortest horizon likely reflects a combination of higher signal freshness and a narrower evaluation window (13 rebalance months). At 12~months, the MV portfolio earns 16.8\% annualised with 7.3\% volatility (SR~$= 2.31$) and a maximum drawdown of 3.7\%. The matched S\&P~500 cap-weighted Sharpe ratio over the same window is~1.31, indicating that the MV portfolio substantially outperforms the matched price-only benchmark on a Sharpe-ratio basis while delivering better downside protection (MDD $-3.7\%$ vs.\ $-9.7\%$); however, the 30-month evaluation window coincides almost entirely with a rising equity market, and the strategy's behaviour in a downturn remains untested.

\begin{table}[htbp]
\centering
\small
\sbox0{%
\begin{tabular}{llrrrrrrrr}
\toprule
& & & \multicolumn{3}{c}{MV-Optimised} & \multicolumn{3}{c}{Equal-Weight} & S\&P 500 \\
\cmidrule(lr){4-6} \cmidrule(lr){7-9}
& & Months & Ann.\ ret & Vol & SR & Ann.\ ret & Vol & SR & SR \\
\midrule
\multicolumn{10}{l}{\textit{Panel A: Aggregate portfolio by signal life}} \\[3pt]
3-month  & & 13 & 21.93 &  6.21 & 3.53 & 25.31 & 16.27 & 1.56 & 2.00 \\
6-month  & & 22 & 17.71 &  7.60 & 2.33 & 17.42 & 14.49 & 1.20 & 1.31 \\
12-month & & 30 & 16.83 &  7.27 & 2.31 & 21.61 & 13.96 & 1.55 & 1.31 \\
\midrule
\multicolumn{10}{l}{\textit{Panel B: Per-model portfolios (12-month signal life)}} \\[3pt]
GPT-4o        & Oct 2023 & 12 &  17.81 &  5.70 &  3.12 &  21.62 & 17.43 &  1.24 & 1.85 \\
GPT-4-Turbo   & Dec 2023 & 12 &  20.15 &  6.80 &  2.96 &  26.37 & 14.15 &  1.87 & 2.20 \\
\addlinespace
GPT-4.1       & Jun 2024 & 12 &  15.14 &  8.17 &  1.85 &   7.93 & 13.10 &  0.61 & 0.50 \\
GPT-4.1-mini  & Jun 2024 & 12 &   7.80 &  7.56 &  1.03 &   7.93 & 13.10 &  0.61 & 0.50 \\
GPT-4.1-nano  & Jun 2024 & 12 &  15.88 &  9.21 &  1.72 &   7.93 & 13.10 &  0.61 & 0.50 \\
o3            & Jun 2024 & 12 &   8.73 &  7.81 &  1.12 &   7.93 & 13.10 &  0.61 & 0.50 \\
GPT-5-mini    & May 2024 & 12 &   7.81 &  8.55 &  0.91 &   7.93 & 13.10 &  0.60 & 0.50 \\
GPT-5-nano    & May 2024 & 12 &   4.29 &  8.58 &  0.50 &   7.93 & 13.10 &  0.61 & 0.50 \\
\addlinespace
GPT-5         & Sep 2024 & 12 &  10.77 &  8.09 &  1.33 &  25.17 & 13.52 &  1.86 & 0.46 \\
GPT-5.2       & Aug 2025 &  7 &  10.85 &  8.47 &  1.28 &  24.73 &  7.74 &  3.21 & 3.02 \\
GPT-5.4       & Aug 2025 &  7 &  18.97 &  7.99 &  2.38 &  24.73 &  7.74 &  3.21 & 3.02 \\
\addlinespace
\textit{Average} & & & 12.56 & 7.90 & 1.66 & 15.47 & 12.58 & 1.37 & 1.23 \\
\bottomrule
\end{tabular}}%
\begin{minipage}{\wd0}
\caption{Portfolio performance: signal-life comparison and per-model results.}
\label{tab:portfolio_summary}
\usebox0
\par\smallskip
\footnotesize
GPT-3.5-Turbo is excluded: its September~2021 training cutoff precedes the next model by over two years, creating a large temporal gap with no overlapping signal coverage and conflating the 2022 bear market with the denser post-2023 sample. Panel~A reports the aggregate portfolio using one model per cutoff date (the most capable available). Panel~B reports individual model portfolios with a 12-month signal life, capped at 12~rebalance months; dates indicate the knowledge-cutoff month; the portfolio becomes active one month later. GPT-5.2 and GPT-5.4 have only 7~months of post-cutoff data available. The S\&P~500~SR column reports the Sharpe ratio of the cap-weighted S\&P~500 Price Index over each row's exact evaluation window, providing a like-for-like benchmark comparison; models sharing a cutoff date share the same benchmark window. The MV-Optimised portfolio uses regularised mean-variance optimisation with the Axioma US4-MH factor risk model ($\lambda = 1$, diagonal shrinkage $\alpha = 0.10$, z-score winsorisation at $\pm 2.5$, long-only, 1.5\% position cap). The Equal-Weight portfolio invests uniformly in the same universe. Ann.\ ret = annualised return (\%); Vol = annualised volatility (\%); SR = Sharpe ratio; all returns gross of transaction costs. Months = number of rebalance months in sample.
\end{minipage}
\end{table}

Panel~B reports the per-model results for a 12-month signal life, capped at 12~rebalance months (GPT-5.2 and GPT-5.4 have 7~months of post-cutoff data). Each checkpoint defines a separate portfolio whose alpha derives exclusively from that model's outlook scores. All~11 models produce positive MV Sharpe ratios, with a cross-model average of~1.66 (annualised return 12.6\%, volatility 7.9\%). The cross-model average MV Sharpe exceeds the average matched S\&P~500 Sharpe of~1.23, and eight of~11 models individually beat their benchmark (one ties), although the advantage is not universal across evaluation windows. The October~2023 and December~2023 cohorts face a strong market (S\&P SR~$= 1.85$ and $2.20$), while the June~2024 cluster benefits from a more moderate price-only benchmark (S\&P SR~$= 0.50$). In the June~2024 cluster, five of six models exceed the S\&P~500 Sharpe and the sixth matches it (GPT-4.1~$= 1.85$, GPT-4.1-nano~$= 1.72$, o3~$= 1.12$, GPT-4.1-mini~$= 1.03$, GPT-5-mini~$= 0.91$, GPT-5-nano~$= 0.50$ vs.\ S\&P~$= 0.50$). GPT-5 likewise outperforms its matched benchmark (1.33 vs.\ 0.46). Within the June~2024 cluster the MV Sharpe broadly follows the capability gradient, providing suggestive portfolio-level support for the model-sophistication finding in Section~\ref{sec:results_h4}.

\begin{table}[htbp]
\centering
\small
\sbox0{%
\begin{tabular}{llr rr rr}
\toprule
& & & \multicolumn{2}{c}{Sharpe ratio} & \multicolumn{2}{c}{Max drawdown (\%)} \\
\cmidrule(lr){4-5} \cmidrule(lr){6-7}
& & Months & MV & S\&P 500 & MV & S\&P 500 \\
\midrule
GPT-4o        & Oct 2023 & 12 &  3.12 &  1.85 &  $-$1.47 &  $-$4.95 \\
GPT-4-Turbo   & Dec 2023 & 12 &  2.96 &  2.20 &  $-$2.05 &  $-$4.95 \\
\addlinespace
GPT-4.1       & Jun 2024 & 12 &  1.85 &  0.50 &  $-$3.58 &  $-$9.66 \\
GPT-4.1-mini  & Jun 2024 & 12 &  1.03 &  0.50 &  $-$3.72 &  $-$9.66 \\
GPT-4.1-nano  & Jun 2024 & 12 &  1.72 &  0.50 &  $-$3.89 &  $-$9.66 \\
o3            & Jun 2024 & 12 &  1.12 &  0.50 &  $-$3.76 &  $-$9.66 \\
GPT-5-mini    & May 2024 & 12 &  0.91 &  0.50 &  $-$4.91 &  $-$9.66 \\
GPT-5-nano    & May 2024 & 12 &  0.50 &  0.50 &  $-$5.82 &  $-$9.66 \\
\addlinespace
GPT-5         & Sep 2024 & 12 &  1.33 &  0.46 &  $-$3.66 &  $-$9.66 \\
GPT-5.2       & Aug 2025 &  7 &  1.28 &  3.02 &  $-$1.87 &  $-$1.04 \\
GPT-5.4       & Aug 2025 &  7 &  2.38 &  3.02 &  $-$1.28 &  $-$1.04 \\
\addlinespace
\textit{Average} & & & 1.66 & 1.23 & $-$3.27 & $-$7.24 \\
\midrule
\multicolumn{3}{l}{Models beating S\&P 500} & 8/11 & & 9/11 & \\
\bottomrule
\end{tabular}}%
\begin{minipage}{\wd0}
\caption{Risk comparison: MV portfolio versus matched S\&P~500 (12-month signal life).}
\label{tab:risk_comparison}
\usebox0
\par\smallskip
\footnotesize
Each row compares the per-model MV-optimised portfolio from Table~\ref{tab:portfolio_summary} with the S\&P~500 cap-weighted Price Index over the identical calendar window. For the Sharpe ratio column, ``beating'' means a higher Sharpe; for max drawdown, it means a smaller (less negative) drawdown. All returns are gross of transaction costs.
\end{minipage}
\end{table}

The cross-model average MV Sharpe ratio of~1.66 exceeds the average matched S\&P~500 Sharpe of~1.23, and 8 of~11 models individually beat their benchmark on this metric (one ties). The risk-management picture is likewise distinctly favourable. Table~\ref{tab:risk_comparison} reports the maximum drawdown of each per-model MV portfolio alongside the matched S\&P~500 cap-weighted index over the same calendar window. Nine of~11 models exhibit a smaller maximum drawdown than the S\&P~500. The cross-model average MV drawdown is $-3.3\%$ versus $-7.2\%$ for the benchmark, while average volatility is 7.9\% versus 11.8\%. These advantages are largest for the June~2024 cluster, where the MV portfolios cap drawdowns at 3.6--5.8\% against the S\&P~500's $-9.7\%$. The two exceptions are GPT-5.2 (MDD $-1.9\%$ vs.\ S\&P $-1.0\%$) and GPT-5.4 (MDD $-1.3\%$ vs.\ S\&P $-1.0\%$), both over the short August~2025 window where the benchmark drawdown is itself minimal. Overall, the backtest indicates stronger sample-period risk-adjusted performance and lower drawdowns relative to the matched price-only benchmark, though the advantage is not universal across windows.

\section{Conclusion}
\label{sec:conclusion}

Frozen LLM checkpoints extract information from pre-cutoff public text that is positively associated with returns across twelve model checkpoints, multiple horizons and a battery of robustness checks, after controlling directly for market-implied cheapness metrics (inverse P/E, implied cost of equity and inverse EV/EBITDA). The coefficient broadly increases with the return horizon, despite a non-monotonic intermediate dip, consistent with gradual price convergence, and appears widespread across firm size and information-environment characteristics. The evidence suggests that the predictive content resides in the LLM's qualitative textual assessment rather than in its interaction with market-implied valuations: the outlook score dominates cheapness measures in a head-to-head decomposition and combining the two via an additive mismatch does not improve on the score alone.

The results are suggestive about possible mechanisms underlying this information gap. In the pooled panel, the outlook-score effect strengthens with analyst and price-target coverage, though single-model evidence is directionally consistent but weaker. This pattern is more naturally aligned with a narrative-congestion interpretation than with a simple limited-attention story, though neither can be ruled out definitively. One interpretation is that the bottleneck is the cost of aggregating dispersed qualitative information across many documents---a task at which LLMs have a comparative advantage---rather than a lack of investor monitoring. This distinction has implications for the durability of the phenomenon: unlike limited-attention frictions, which diminish as more investors adopt AI-based tools, narrative congestion may persist or intensify as the volume of corporate disclosure continues to grow.

We do not claim a formal rejection of market efficiency. The joint-hypothesis problem applies here as elsewhere: the outlook score may partly capture compensation for a risk dimension not spanned by our factor controls. With only six unique cutoff dates, a formal time-series alpha test is not feasible, and the portfolio's factor exposures indicate substantial overlap with known risk premia (profitability and investment). Importantly, the core result survives corrections for overlapping observations: Driscoll--Kraay standard errors on the full panel ($t = 5.43$), a clean Fama--MacBeth procedure using one model per cutoff date with Newey--West adjustment ($t = 2.25$, bootstrap $p = 0.019$) and the horizon structure under both approaches consistently confirm cross-sectional predictability. What we can say more cautiously is that the LLM's reading of the pre-cutoff textual record contains information not captured by our standard contemporaneous valuation controls and that this information gap is associated with statistically robust cross-sectional return predictability after controlling directly for these factors. Taken together, the combination of fundamental predictability (H1), horizon convergence (H4) and the suggestive narrative-congestion pattern (H3) points toward an information-processing friction rather than a systematic risk premium, but the evidence does not permit a definitive conclusion.

Several limitations warrant further acknowledgement. First, all checkpoints come from a single provider (OpenAI) and the signal may partly reflect idiosyncratic properties of that model family rather than a universal feature of LLM-derived information. Replication with checkpoints from other providers (e.g., Anthropic, Google, Meta) would strengthen the external validity of the findings. Second, the post-cutoff window for the most recent models (August~2025 cutoff) spans only seven months, limiting the horizon analysis for the frontier tier. As time passes and longer return histories accumulate, more precise estimates of long-horizon convergence will become available. Third, while both the twelve-model and the more conservative six-cutoff Fama--MacBeth procedures confirm cross-sectional significance, the twelve models map to only six unique calendar months, precluding a formal time-series alpha test. A longer panel of checkpoints spanning more calendar periods would enable proper risk-adjusted return estimation. Fourth, the outlook score is constructed from a single prompt design; alternative elicitation strategies---different scoring scales, multi-turn dialogues, or chain-of-thought prompting for standard models---may extract different facets of the LLM's embedded knowledge.

For practitioners, the findings suggest that LLM-derived fundamental assessments contain economically significant information not yet captured by standard valuation metrics. A preliminary portfolio application in Section~\ref{sec:portfolio} indicates that the cross-sectional signal can be translated into an implementable strategy with meaningful risk-management benefits: over a 30-month evaluation window, the aggregate MV portfolio achieves a gross-of-cost Sharpe ratio of~2.31 with a 3.7\% maximum drawdown, exceeding the S\&P~500 Price Index on a price-return basis (SR~$= 1.31$) and delivering substantially better downside protection (MDD $-3.7\%$ vs.\ $-9.7\%$). Per-model benchmarking against the S\&P~500 over matched calendar windows is consistent with this advantage: eight of~11 models exceed the benchmark on Sharpe ratio (average 1.66 vs.\ 1.23) and nine produce smaller drawdowns (average $-3.3\%$ vs.\ $-7.2\%$). The signal's relative edge appears clearest in moderate-market environments---in the June~2024 cluster, five of six models surpass the matched S\&P~500 Sharpe of~0.50 (the sixth matches it), while all six achieve lower drawdowns---and the October~2023 cohort faces a stronger market that is harder to beat on a risk-adjusted basis. The factor loadings indicate that the strategy tilts toward profitable, capital-light firms---a profile consistent with quality investing, though the cross-sectional outlook-score coefficient survives direct controls for these factors. The durability of this edge depends on the speed of adoption: as market participants increasingly deploy similar tools, the informational advantage of any single LLM checkpoint should erode, consistent with the anomaly-attenuation dynamics documented by \cite{mclean2016does}. The cross-model comparison already hints at this trajectory, though the current sample is too small for definitive conclusions.

Several extensions are natural: expanding the checkpoint universe to multiple providers and model generations for richer capability--signal and model-age-decay estimates; applying the framework to international equity markets to test geographic generality; decomposing the outlook score by textual source to identify which components of the public record are most underprocessed; and extending the time-capsule methodology to fixed-income, commodity and macroeconomic forecasting settings.

\newpage
\appendix
\label{app:appendix}

\section{LLM Prompt Template}
\label{app:prompt}

\begin{verbatim}
System Prompt:
You are an unbiased CFA-level equity analyst. Assume 
"today" is {CUT_OFF_DATE}. You must IGNORE any event,
price, or fact that occurred after that date. Base every
judgment solely on information publicly available on or
before {CUT_OFF_DATE} and the identifiers provided 
for the company below. Respond in English.

User Prompt:
Using only pre-{CUT_OFF_DATE} knowledge, produce a
12-month forward **business outlook** for {COMPANY_NAME}
(Ticker: {TICKER}, ISIN: {ISIN}) relative to peers in
its industry. Focus exclusively on business 
fundamentals—revenue growth, profitability trends and
key operational risks. DO NOT use or reference valuation
metrics, market prices, or technical indicators.

Your response must follow the JSON schema below **exactly**.
Do not include any additional commentary, formatting, or code
fences. If uncertain, lower confidence and avoid speculation.

OUTPUT FORMAT (strict JSON)
{
  "knowledge_cutoff_date": "{CUT_OFF_DATE}",
  "firm": {
    "name": "{COMPANY_NAME}",
    "ticker": "{TICKER}",
    "country": "{COUNTRY}",
    "industry_code": "{INDUSTRY_CODE}"
  },
  "horizon_months": 12,
  "scores": {
    "outlook": integer [-10..10],
    "growth": integer [-10..10],
    "profitability": integer [-10..10],
    "risk": integer [-10..10],
    "confidence": integer [0..100]
  },
  "distributions": {
    "revenue_growth_bins_pct": {
      "less_than_minus_10": int,
      "minus_10_to_0": int,
      "plus_0_to_5": int,
      "plus_5_to_10": int,
      "greater_than_plus_10": int
    },
    "eps_growth_bins_pct": {
      "less_than_minus_20": int,
      "minus_20_to_0": int,
      "plus_0_to_10": int,
      "plus_10_to_25": int,
      "greater_than_plus_25": int
    },
    "margin_change_bins_pct": {
      "less_than_minus_2pp": int,
      "minus_2pp_to_0pp": int,
      "plus_0pp_to_1pp": int,
      "plus_1pp_to_2pp": int,
      "greater_than_plus_2pp": int
    }
  },
  "drivers": [
    2-5 tags from this set:
    ["product_cycle", "pricing_power", "regulation", "competition",
     "supply_chain", "capital_needs", "macro_exposure", "IP_legal",
     "mgmt_execution", "network_effects", "customer_concentration",
     "input_costs"]
  ],
  "rationale_short": "concise summary < 30 tokens; cite 1-3 key drivers;
                       no facts after {CUT_OFF_DATE}",
  "knowledge_coverage": integer [0..100]
}

HARD CONSTRAINTS
- Return only a valid JSON object (no code fences, markdown, or extra text).
- All numeric fields must be numbers, not strings.
- Each sub-distribution must sum to exactly 100.
- Do not include placeholders (e.g., __REQUIRED__, TODO, ...).
- Lower confidence if the company or industry knowledge is limited.
\end{verbatim}


\end{document}